\documentclass{aa}  
\usepackage{graphicx}
\usepackage{txfonts}
\usepackage{natbib}
\usepackage{color}
\bibpunct{(}{)}{;}{a}{}{,}

\begin{document}
   \title{Comparison of synthetic maps from truncated jet-formation 
          models with YSO jet observations}
   \subtitle{II. The effect of varying inclinations}

\author{Matthias Stute\inst{1}
  \and
  Jos\'e Gracia\inst{2}
}

\offprints{Matthias Stute,\\ \email{matthias.stute@tat.physik.uni-tuebingen.de}}

\institute{
  Institute of Astronomy and Astrophysics, Section Computational Physics,
  Eberhard Karls Universit\"at T\"ubingen, Auf der Morgenstelle 10, 72076 
  T\"ubingen, Germany
  \and
  High Performance Computing Center Stuttgart (HLRS), Universit\"at 
  Stuttgart, 70550 Stuttgart, Germany
}

   \date{Received ; accepted }

  \abstract
   {Analytical radially self-similar models are the best available solutions 
    describing disk-winds but need several improvements. In a previous article, 
    we introduced models of jets from truncated disks, i.e. evolved in time 
    numerical simulations based on a radially self-similar MHD solution but 
    including the effects of a finite radius of the jet-emitting disk and thus 
    the outflow. In paper I of this series, we compared these models with 
    available observational data varying the jet density and velocity, the mass 
    of the protostar and the radius of the aforementioned truncation.}
   {In paper I, we assumed that the jet lies in the plane of the sky. In this 
     paper, we investigate the effect of different inclinations of the jet.}
   {In order to compare our models with observed jet widths inferred from
    recent optical images taken with HST and AO, we create again emission maps 
    in different forbidden lines and from such emission maps, we determine the 
    jet width as the full-width half-maximum of the emission.}
   {We can reproduce the jet width of DG Tau and its variations very well and 
     the derived inclination of 40$^\circ$ is in excellent agreement with 
     literature values of 32--52$^\circ$. In CW Tau we overestimate the 
     inclination in our best-fit model. In the other objects, we cannot find 
     appropriate models which reproduce the variations of the observed jet 
     widths, only the average jet width itself is well modeled as in paper I.}
   {We conclude that truncation -- i.e. taking into account the finite radius 
     of the jet launching region -- is necessary to reproduce the observed jet 
     widths and our simulations limit the possible range of truncation radii. 
     The effects of inclination are important for modeling the intrinsic 
     variations seen in observed jet widths. Our models can be used to infer 
     independently the inclinations in the observed sample, however, a 
     parameter study with a finer grid of parameters is needed.}


   \maketitle

\section{Introduction}

Astrophysical jets and disks \citep{Liv09} seem to be inter-related, notably in 
the case of young stellar objects (YSOs), wherein jet signatures are well 
correlated with the infrared excess and accretion rate of the circumstellar 
disk \citep{CES90, HEP04}. Disks provide the plasma which is outflowing in the 
jets, while jets in turn provide the disk with the needed angular momentum 
removal in order that accretion in the protostellar object takes place 
\citep{Har09}. On the theoretical front, the most widely accepted description 
of this accretion-ejection phenomenon \citep{Fer07} is based on the interaction 
of a large scale magnetic field with an accretion disk around the central 
object. Then, plasma is channeled and magneto-centrifugally accelerated along 
the open magnetic field lines threading the accretion disk, as firstly described
in \citet{BlP82}. Several works have extended this study either by semi-analytic
models using radially self-similar solutions of the full magnetohydrodynamics 
(MHD) equations with the disk treated as a boundary condition \citep{VlT98}, 
or, by selfconsistently treating numerically the disk-jet system 
\citep[e.g.][]{ZFR07}.

The original \citet{BlP82} model, however, has serious limitations for a needed 
meaningful comparison of its predictions with observations. Singularities exist 
at the jet axis, the outflow is not asymptotically superfast, and most 
importantly, the disk has no intrinsic scale with the result that the jet 
formally extends to radial infinity, to mention just a few. The singularity at 
the axis can be easily taken care of by numerical simulations extending the 
analytical solutions close to this symmetry axis 
\citep[][GVT06 hereafter]{GVT06}. The outflow speed at large distances may be 
tuned to cross the corresponding limiting characteristic, with the result that 
the terminal wind solution is causally disconnected from the disk and hence 
perturbations downstream of the superfast transition (as modified by 
self-similarity) cannot affect the whole structure of the steady disk-wind 
outflow \citep[][V00 hereafter]{VTS00} which has also been shown to be 
structurally stable \citep[][M08 hereafter]{MTV08}. The next step of 
introducing a scale in the disk has been undertaken in a previous paper 
\citep{STV08}, wherein we presented numerical simulations of truncated flows 
whose initial conditions are based on analytical self-similar models.

In order to test our truncated models, we then applied our simulations to 
observations \citep[][paper I hereafter]{SGT10}. In recent years, many NIR and 
optical data have become available exploring the morphology and kinematics of 
the jet launching region 
\citep[e.g.][and references therein]{DCL00,RDB07,Dou08}. HST and adaptive
optics observations give access to the innermost regions of the wind, where the 
acceleration and collimation occurs \citep{RMD96,DCL00,WRB02,HEP04}. Since YSO 
jets emit in a number of atomic (and molecular) lines, we use a synthetic 
emission code to create emission maps in different forbidden lines which were 
used by other authors to extract the jet width from images. In paper I, we 
compared the observed jet widths with those extracted from our synthetic images.
A similar study has been done by \citet{Fer97,CaF00,GCF01,DCF04}, however, 
using untruncated disk wind models.

In paper I, however, we always assumed the inclination of the jet to be 
90$^\circ$. The influence of varying inclinations on the derived synthetic jet 
widths will be examined now in this paper.

The remainder of the paper is organized as follows: we briefly review the 
initial set-up of the numerical simulations and describe our procedure for the 
comparison with observations in Sec. \ref{sec_models}. The results of our 
studies are presented in Sec. \ref{sec_results}. In Sec. \ref{sec_bestfit} we 
describe our best-fit models for each object in the sample and compare derived 
inclinations with literature values. Finally, we conclude with the 
implications of the results in terms of the structure of the disk and the 
respective launching radii of the jets in YSOs. In appendix \ref{sec_all_lw}, 
we present the extracted line widths for all models, runs and inclinations for 
the sake of completeness. Jet velocities derived from synthetic position-
velocity diagrams are given in Appendix \ref{sec_pv_vel}.

\section{The models} \label{sec_models}

\subsection{Initial set-up and numerical simulations}

This work is based on the results of our numerical simulations discussed in 
\citet{STV08} and two new models already added in paper I (Table 
\ref{tbl_num_models}). We solved the MHD equations with the PLUTO 
code\footnote{http://plutocode.to.astro.it/} \citep{MBM07} starting from 
initial conditions set according to a steady, radially self-similar solution 
as described in V00 which crosses all three critical surfaces. At the symmetry 
axis, the analytical solution was modified as described in GVT06 and M08.

To study the influence of the truncation of the analytical solution, we divide
our computational domain up into a jet region and an external region, separated 
by a truncation field line $\alpha_{\rm trunc}$. For lower values of the 
normalized magnetic flux function, i.e. $\alpha < \alpha_{\rm trunc}$ -- or 
conversely smaller cylindrical radii -- our initial conditions are fully 
determined by the solution of V00 and the modification of GVT06 and M08 close to
the axis. In the outer region, we modify all quantities and initialize them with
another analytical solution but with modified parameters (models with outer 
truncation; SC1a--g, SC2, SC4). In models SC1a--g we varied the truncation 
radius and within models SC1a, SC2 and SC4, we probed different modifications in
the outer region. Two simulations have been performed with inner truncation 
(SC3, SC5), i.e. the analytical solution with modified quantities is inside the 
original analytical solution. For further details, we refer the reader to 
\citet{STV08}.

\begin{table*}
\caption{List of numerical models}
\label{tbl_num_models}
\centering
\begin{tabular}{lll}
\hline\hline
Name & $R_{\rm trunc}$ $[ R_0 ]$ & Description   \\
\hline
model ADO  & $\infty$ & unchanged analytical solution of V00 \\
model SC1a & 5.375    & analytical solution is truncated for $R > R_{\rm trunc}$ 
(outer truncation), quantities are scaled down by factors of $10^{-6\ldots-1.5}$
\\
model SC1b & 5.125    & same as model SC1a, but different $R_{\rm trunc}$ \\
model SC1c & 4.875    & same as model SC1a, but different $R_{\rm trunc}$ \\
model SC1d & 3.625    & same as model SC1a, but different $R_{\rm trunc}$ \\ 
model SC1e & 2.625    & same as model SC1a, but different $R_{\rm trunc}$ \\ 
model SC1f & 2.375    & same as model SC1a, but different $R_{\rm trunc}$ \\ 
model SC1g & 0.575    & same as model SC1a, but different $R_{\rm trunc}$ \\ 
model SC2  & 5.375    & analytical solution is truncated for $R > R_{\rm trunc}$ 
(outer truncation), scale factors of $10^{-2\ldots-1}$, density unchanged \\ 
model SC3  & 5.375    & same as model SC2, but analytical solution is truncated 
for $R < R_{\rm trunc}$ (inner truncation) \\
model SC4  & 5.375    & analytical solution is truncated for $R > R_{\rm trunc}$ 
(outer truncation), scale factors of $10^{-2\ldots-1}$, velocity unchanged \\ 
\hline
\end{tabular}
\noindent
For further details, we refer the reader to \citet{STV08}. The meaning of $R_0$
is described in Sec. \ref{sec_norm}.
\end{table*}

\subsection{Synthetic emission maps and jet width extraction}

We use the synthetic emission code OpenSESAMe v0.1 to produce synthetic 
observations from our simulations in different consecutive stages. The first 
stage approximates the chemical composition of the plasma by locally solving a 
chemical network under the assumption of stationarity. The second stage 
calculates the statistical equilibrium of level populations for each ion of 
interest as a function of temperature and density and yields the emissivity for 
individual transitions of interest. Further stages take care of integration 
along the line-of-sight and projection. Finally, the ideal maps are convolved 
with a Gaussian point-spread-function (PSF) to mimic the finite spatial 
resolution of a given instrument. These synthetic emission maps are then 
quantitatively analysed with similar techniques used on real observed maps.

\begin{figure}[!htb]
  \centering
  \includegraphics[width=\columnwidth]{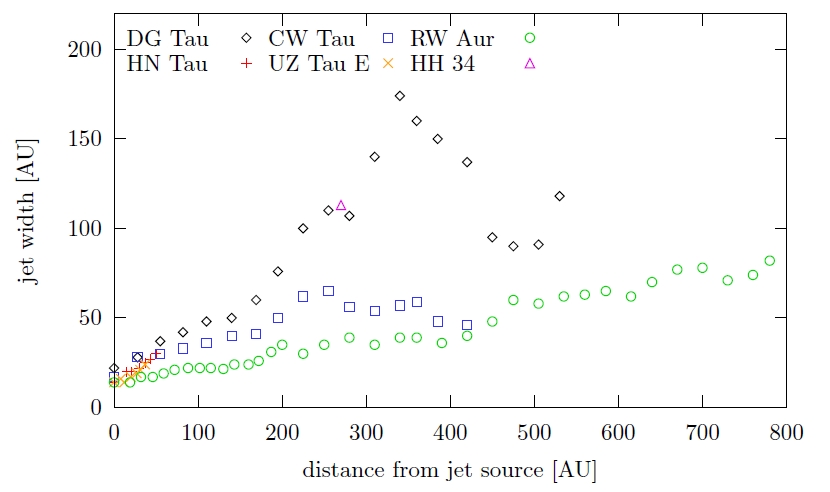}
  \caption{Variation of jet width (FWHM) derived from [SII] and [OI] images as 
    a function of distance from the source. Data points are from CFHT/PUEO and 
    HST/STIS observations of DG Tau (diamonds), HN Tau (plus signs), CW Tau 
    (squares), UZ Tau E (crosses), RW Aur (circles), HH 34 (one triangle); data 
    are taken from \citet{RDB07} for distances below 200 AU and \citet{DCL00} 
    beyond this distance.}
  \label{Fig_observations}
\end{figure}

We compare the width of jets measured from HST and AO observations 
\citep[][and references therein]{DCL00,RDB07,Dou08} to synthetic emission maps
calculated from our MHD models. We convolved the maps with a Gaussian with a 
FHWM of 15 AU ($\sigma = 6.37$ AU) throughout this paper, equivalent to HST's 
resolution of 0.1'' at a distance of 150 pc. We use a sample of eight jets: DG 
Tau, HN Tau, CW Tau, UZ Tau E, RW Aur, HH34, HH30 and HL Tau (Fig. 
\ref{Fig_observations}). In order to determine the width of the jets in our 
models, we use a method which is as close as possible to that applied by the 
observers. We create convolved synthetic maps for the emission in the [SII] 
$\lambda$6731 and [OI] $\lambda$6300 lines for each numerical model and each 
run of OpenSESAMe and determine the jet width from the map's full-width 
half-maximum (FWHM) as a function of distance along the axis. 

\begin{figure*}[!htb]
  \centering
  \includegraphics[width=\textwidth]{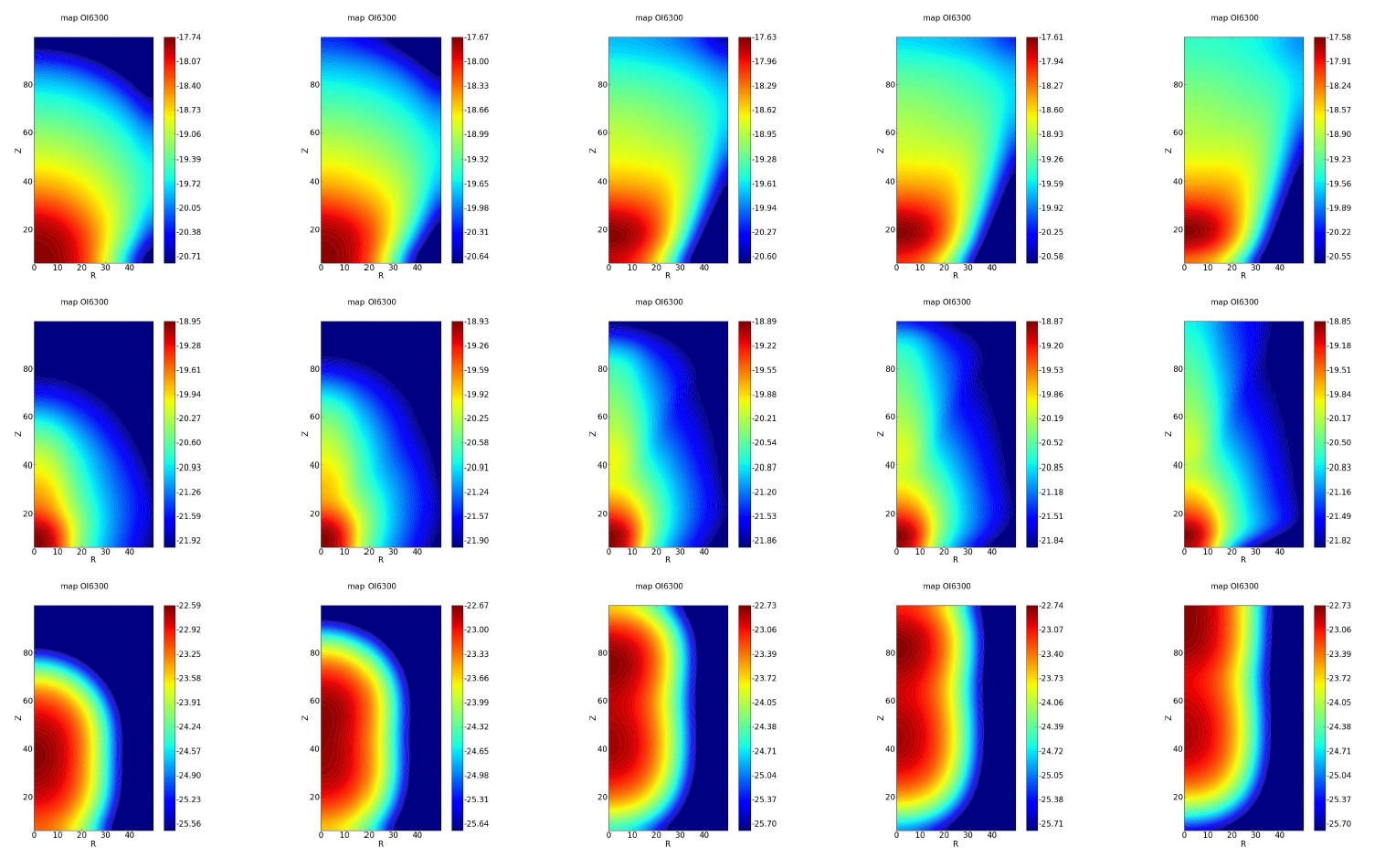}
  \caption{Synthetic emission maps of the [OI] $\lambda$6300 line, convolved 
    with a Gaussian PSF with a FWHM of 15 AU, for run (500,1000,0.5) and the 
    untruncated model ADO (top) and the model SC1a with outer truncation 
    (middle) and for run (500,100,0.2) and model SC3 with inner truncation 
    (bottom). The inclinations are 30$^\circ$, 40$^\circ$, 60$^\circ$, 70$^\circ$ 
    and 90$^\circ$ (from left to right). The flux is in units of erg
    s$^{-1}$ cm$^{-2}$.}
  \label{Fig_emissmaps}
\end{figure*}

\begin{figure*}[!htb]
  \centering
  \includegraphics[width=\textwidth]{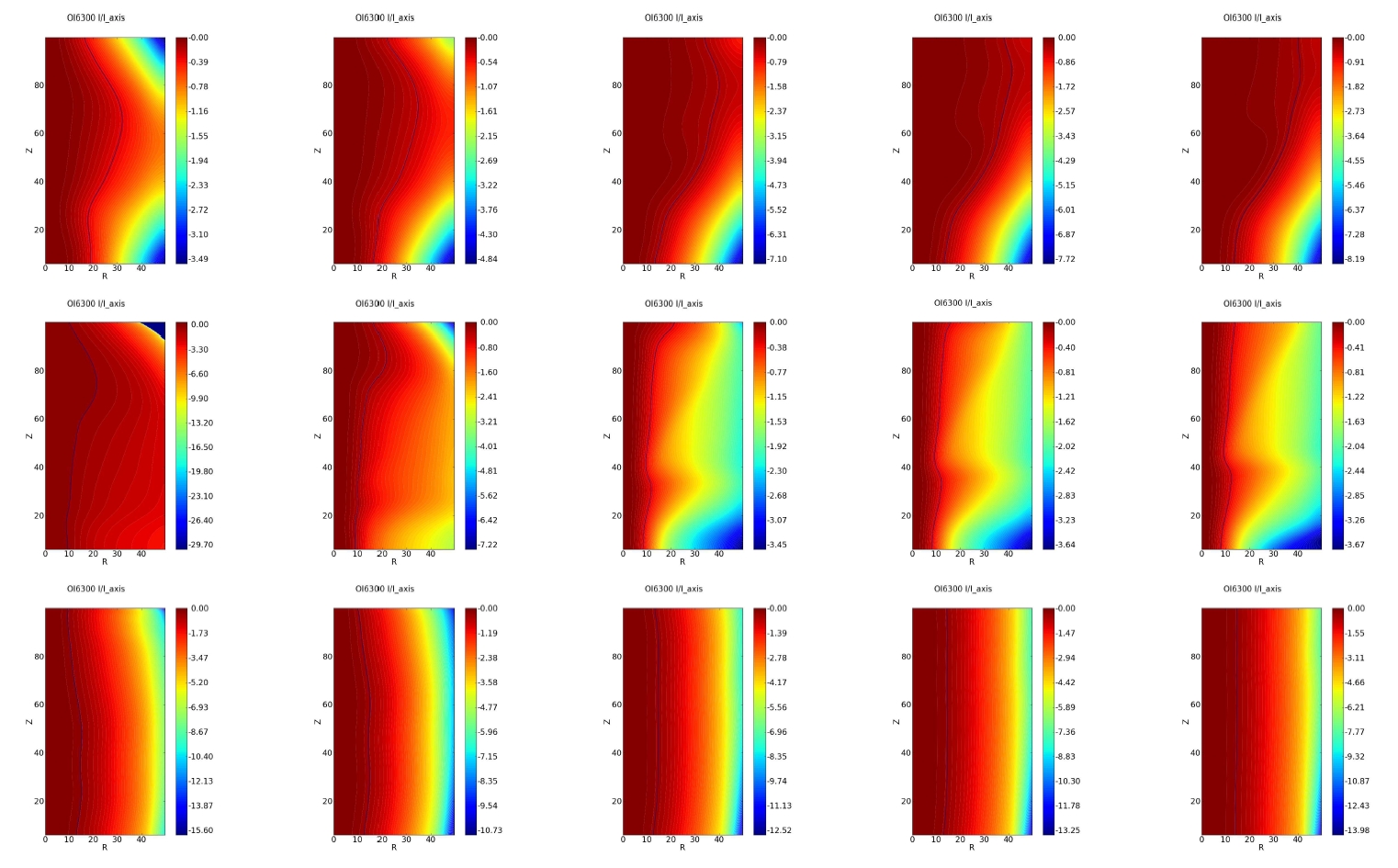}
  \caption{same as Fig. \ref{Fig_emissmaps}, however, normalized by the 
    intensity on the jet axis for each row; also plotted is the contour line 
    where $I/I_{\rm axis} = 0.5$, i.e. the position of the FWHM.}
  \label{Fig_emissmaps_normalized}
\end{figure*}

\subsection{Normalizations} \label{sec_norm}

In order to compare our results with observations, i.e. in order to run
OpenSESAMe correctly, we have to scale the dimensionless quantities in which 
PLUTO performs its calculations \citep{STV08} to physical units by providing 
scaling factors for density $\rho_0$, pressure $p_0$, velocity $v_0$, magnetic 
field strength $B_0$, a length scale $R_0$ and a mass scale $\mathcal{M}$. 
However, in terms of the normalizations used in the PLUTO code, only three of 
those quantities are indepedent. A possible choice is the mass of the central 
object, velocity scale and density scale, while the remaining factors are 
calculated from these. The PLUTO domain is set from (0,6) to 
(50,100) and the physical scale of the jet in AU is simply the PLUTO domain 
multiplied with a length scale $R_0$. In the solution of V00, the length scale 
$R_0$ is connected to the mass of the central object and the velocity 
normalization via
\begin{equation}
\label{length_scale}
R_0 = \frac{\mathcal{G}\,\mathcal{M}}{4\,\,v_0^2} = 
110.9\,\textrm{AU}\,\left(\frac{v_0}{
\textrm{km s$^{-1}$}}\right)^{-2}\,\left(\frac{\mathcal{M}}{
0.5\,\textrm{M}_{\odot}}\right) \,.
\end{equation}

From the velocity and density normalization directly follow the 
normalizations for the magnetic field and pressure as
\begin{eqnarray}
\label{pressure_scale}
p_0 &=& \rho_0\,v_0^2  = 10^{-11}\,\textrm{g cm$^{-1}$ s$^{-2}$}\,
\left(\frac{\rho_0}{10^{-21}\,\textrm{g cm$^{-3}$}}\right) \,
\left(\frac{v_0}{\textrm{km s$^{-1}$}}\right)^2 \,, \\
\label{mag_scale}
B_0 &=& \sqrt{4\,\pi\,\rho_0\,v_0^2} = 11.21\,\textrm{$\mu$G}\,
\left(\frac{\rho_0}{10^{-21}\,\textrm{g cm$^{-3}$}}\right)^{1/2} \,
\left(\frac{v_0}{\textrm{km s$^{-1}$}}\right) \,.
\end{eqnarray}

The mass of the central object affects only the length scale $R_0$. The 
pressure and temperature of the jet, and thus the synthetic emission maps, are 
affected only by the density and velocity scales $\rho_0$ and $v_0$.

As typical jet velocities in YSOs, we assumed in paper I values of 100, 300, 
600 and 1000 km s$^{-1}$, as typical masses of T Tauri stars 0.2, 0.5 and 0.8 
M$_{\odot}$ \citep{HEG95}, and as jet number densities values of 125, 500, 1000 
and $5\times10^4$ cm$^{-3}$. We adopt the nomenclature for our runs as e.g. 
(500,600,0.5) with $n_{\rm jet}$ in cm$^{-3}$, $v_{\rm jet}$ in km s$^{-1}$ and 
$M$ in M$_{\odot}$. 

For numerical reasons, the density shows a steap increase close to the jet axis.
This artefact had to be corrected, therefore we applied the following 
corrections before running OpenSESAMe, as in paper I. We limit the temperature 
to $10^4$ K, the density around the axis by setting the density inside 
1 FWHM to its value at 1 FHWM for each $z$ and the fraction 
$x_{\rm e} = n_{\rm e} / n_{\rm tot}$ to 0.1. These values of $x_{\rm e}$ and $T$ are 
typical of those deduced from analyses of line ratios in jets 
\citep[e.g.][]{BE99, LCD00}. 

With the same requirement that the FWHM of the Gaussian of 15 AU is sampled by 
a reasonable number of pixels (see paper I), the only valid runs are 
\begin{itemize}
\item ($\cdots$,600,0.2), ($\cdots$,600,0.5), ($\cdots$,1000,0.5), 
($\cdots$,1000,0.8) for models ADO, SC1a--b, SC2, SC4
\item ($\cdots$,600,0.2), ($\cdots$,600,0.5), ($\cdots$,1000,0.2), 
($\cdots$,1000,0.5), ($\cdots$,1000,0.8) for models SC1c--f
\item ($\cdots$,600,0.2), ($\cdots$,1000,0.2), ($\cdots$,1000,0.5), 
($\cdots$,1000,0.8) for model SC1g
\item ($\cdots$,100,0.2) for model SC3
\item no valid runs for model SC5.
\end{itemize}

At first sight, the required velocities of 600 and 1000 km s$^{-1}$ in our 
models with outer truncation SC1a--g, SC2 and SC4 seem to be chosen too high in 
comparison with observed values in typical YSO jets. We have to note that the 
velocity $v_{\rm jet}$ is a formal parameter of the model and corresponds to a 
physical velocity component only very close to the source, i.e. is not the 
asymptotic velocity of the flow. 

\begin{figure}[!htb]
  \centering
  \includegraphics[width=\columnwidth]{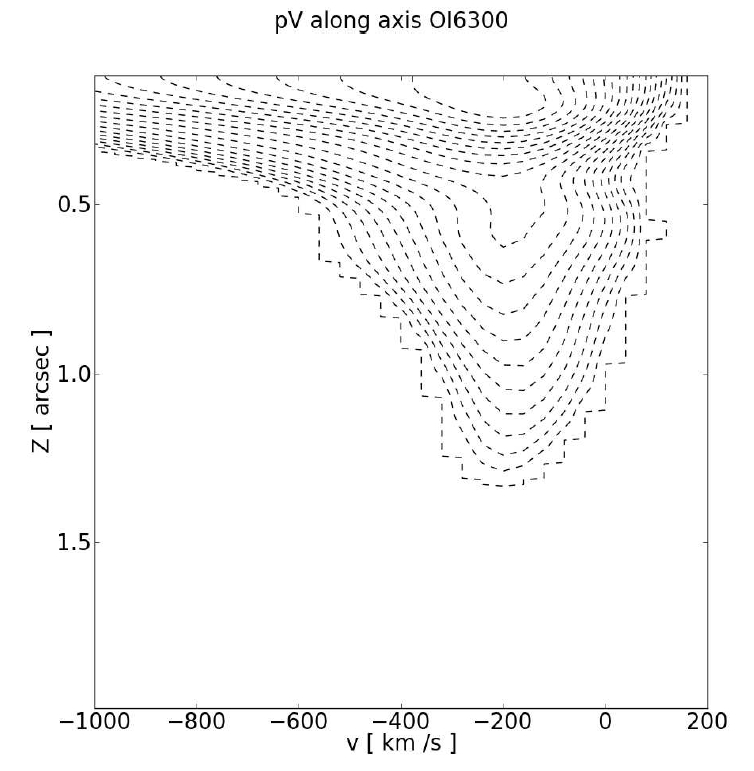} \\
  \includegraphics[width=\columnwidth]{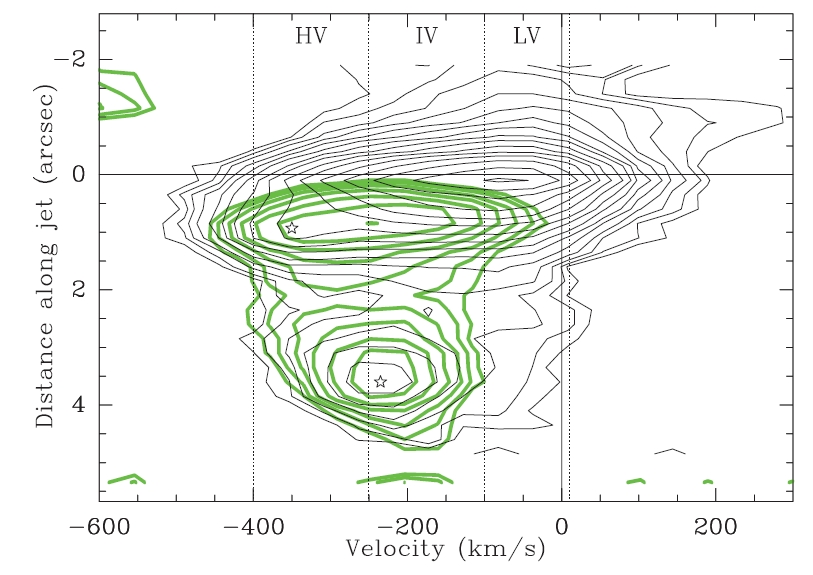}
  \caption{Top: synthetic position-velocity diagram of the [OI] $\lambda$6300 
    line for model SC1a, run (500,1000,0.8) and an inclination of 40$^\circ$, 
    where we assumed a distance to the jet of 150 pc; bottom: observed 
    position-velocity map of DG Tau for [OI] $\lambda$6300 (black thin lines) 
    and [NII] $\lambda$6583 (green thick lines) from \citet{LCD00}. The 
    dynamical range is similar in both plots. The velocity of the jet at 
    distances larger than 0.5'' in the model is 220 km s$^{-1}$ and thus in the 
    range of the observed value of 280 km s$^{-1}$.}
  \label{Fig_pv}
\end{figure}

The observed velocities of YSO jets are usually inferred from either proper 
motions of jet knots (i.e. pattern speeds, not flow speeds) or 
position-velocity diagrams which show the velocities along the line of sight 
convolved with emissivity. In our simulations, proper motions cannot be 
measured due to missing patterns. In Fig. \ref{Fig_pv}, we show a synthetic 
position-velocity diagram for model SC1a, run (500,1000,0.8) and an inclination 
of 40$^\circ$ -- which will be our best-fit model for DG Tau -- and the observed 
map of DG Tau from \citet{LCD00}. We chose a velocity resolution as in the 
observations, also the dynamical range is similar in both plots. We can 
directly see that large velocities around $v_{\rm jet}$ are detectable only close
to the jet source, the velocity of the jet at distances larger than 0.5'' 
is 220 km s$^{-1}$ and thus in the range of the observed value of 280 km 
s$^{-1}$. In Appendix \ref{sec_pv_vel}, we list the velocities of the jet for 
most of our models derived from the synthetic position-velocity diagrams.

\section{Results} \label{sec_results}

Naturally the inclination plays a crucial role for the way how the jet is seen
in synthetic emission maps. This is even more pronounced in the truncated jet 
solutions (Fig. \ref{Fig_emissmaps}). 

\subsection{Synthetic jet widths}

We extracted the jet width from a {\em ratio} of intensities by using the FWHM, 
i.e. we divided the maps by the intensity on the axis and checked where the 
ratio is 0.5. Since in all models we have densities below the critical density
the emissivity is proportional to $\rho^2$ in the whole domain, therefore the 
factor $\rho^2$ should cancel out. In paper I, we already found this behavior, 
the absolute value of the jet density is not important for the extracted jet 
widths. The same is present in this study, thus we will show mainly results 
with a jet density of 500 cm$^{-3}$ throughout this paper.

A second result is the fact that the setup of the external solution is not 
important, i.e. the extracted jet widths of models SC1a, SC2 and SC4 are 
very similar. The only important parameters for our study are indeed the 
truncation radius and the inclination, when the jet velocity and mass of the 
central object are fixed by observations.

In paper I, we found that all untruncated ADO models give too large jet
widths compared to the observations of T Tauri jets. This result is not changed
when varying the inclination. We again conclude that we need an additional 
effect which reduces the derived jet width, in order to be able to reproduce 
all jets in our sample.

In paper I, we already found that the extracted jet width does not follow any 
directly apparent feature in neither the density or temperature map nor the 
emissivity map. A property of the extracted jet width, which was and still is 
intrinsic in all our runs, is the overall structure with at least one maximum. 
The position of maxima is generally only obvious in the normalized emission 
maps, i.e. after dividing by the intensity on the jet axis (Fig. 
\ref{Fig_emissmaps_normalized}). In paper I, we ignored this maximum and 
compared the synthetic jet widths with the observed ones only for large 
distances along the jet axis. Since the observed jet widths, however, also show 
a bumpy profile, it may be worth to further investigate the origin of the maxima
in our simulations and to take them into account for fitting the observations. 

The distance of the first maximum from the jet source increases for increasing 
mass of the central object and decreasing velocity of the jet. It also increases
for increasing degree of truncation. In several models, a second maximum appears
whose distance decreases for increasing degree of truncation. 

\begin{figure}[!htb]
  \centering
  \includegraphics[width=\columnwidth]{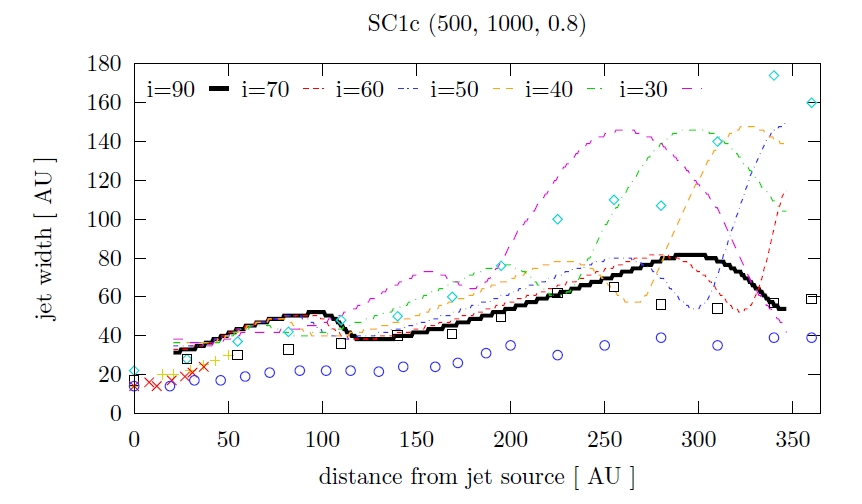}
  \caption{Jet widths in AU derived from synthetic [OI] images as a 
    function of distance from the source in model SC1c and for run 
    (500,1000,0.8); overlaid are the data points of Fig. 
    \ref{Fig_observations}.}
  \label{Fig_inclination_SC1c}
\end{figure}

\begin{figure}[!htb]
  \centering
  \includegraphics[width=\columnwidth]{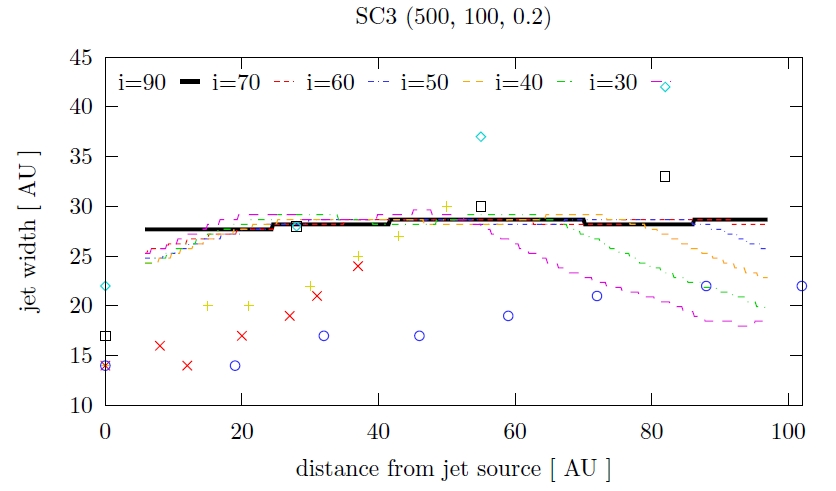}
  \caption{Jet widths in AU derived from synthetic [OI] images as a 
    function of distance from the source in model SC3 and for run (500,100,0.2);
    overlaid are the data points of Fig. \ref{Fig_observations}.}
  \label{Fig_all_inclination_SC3}
\end{figure}

In the model SC3 with inner truncation, the derived jet width is almost 
constant. The effects of varying inclination are therefore only marginal when 
ignoring the geometrical effect mentioned above (Fig. 
\ref{Fig_all_inclination_SC3}).

\subsection{On the origin of variations in the extracted jet widths} \label{sec_variations}

The fact that maxima were already present in our runs in paper I, i.e. in those
with an inclination of 90$^\circ$, shows that we have to differentiate between
intrinsic variations and variations due to geometrical effects. First, we have
to note again that maxima in the extracted jet width are not necessarily 
connected with e.g. density knots. In \citet{STV08}, we found that the model 
with inner truncation SC3 collapses toward the jet axis leading to knots in the 
density maps. In paper I and also in Fig. \ref{Fig_emissmaps} (bottom), we find 
in the corresponding synthetic emission map in [OI] two areas of enhanced 
emission whose position, however, is not the same as that of the density knots. 
Furthermore the extracted jet width is almost constant despite of these areas. 

In order to find an physical origin of the variations, it is useful not to focus
on the first maximum but on the minimum directly behind it. One common feature 
in all our MHD simulations is the fast magnetosonic separatrix surface 
\citep[FMSS,][]{STV08}. This surface is a weak shock causally disconnecting the 
sub-fast flow from the super-fast one. Its position in $Z$ direction 
increases monotonically with increasing $R$ value. Both density and pressure
along the flow line show a jump at the FMSS. As mentioned above, we convolve 
our map with a Gaussian of 15 AU and modify the density inside 15 AU from the 
jet axis. We find that the minimum in the extracted jet width is at the same 
position, where the FMSS has a radius of 15 AU and enters this region. For 
inclinations different to 90$^\circ$, the two crossing points of FMSS with the 
Gaussian both in front of the jet and behind the jet are visible as local 
minima in the extracted jet widths.

\begin{figure}[!htb]
  \centering
  \includegraphics[width=\columnwidth]{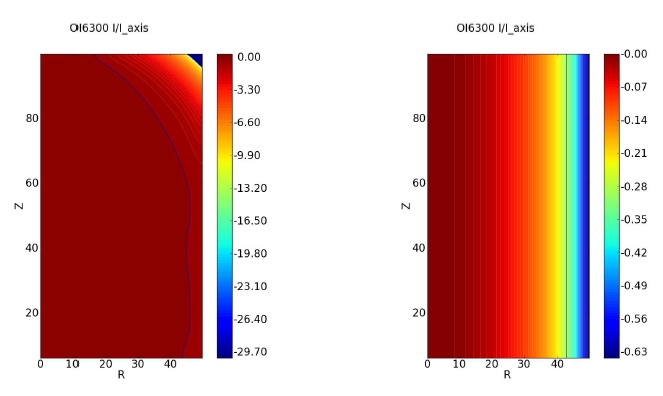}
  \caption{Normalized emission map of the [OI] $\lambda$6300 line for a 
    computational domain with constant density and temperature, the inclination 
    is 40$^\circ$ (left) and 90$^\circ$ (right); also plotted is the contour line 
    where $I/I_{\rm axis} = 0.5$, i.e. the position of the FWHM.}
  \label{Fig_constant_box}
\end{figure}

Beyond physical origins of the maxima, another origin is the geometry of the 
system, i.e. the finiteness of our computational domain. In order to quantify 
this aspect, we calculated the emission maps and extracted jet width for a 
domain with constant density and temperature. Thus also the emission is 
constant across the domain. If seen with an inclination of 90$^\circ$, the 
extracted jet width is constant as expected; if the inclination is 40$^\circ$, 
however, the extracted jet width is almost constant only between about 15 and 68
$R_0$ (Fig. \ref{Fig_constant_box}). These values are dependent on the aspect 
ratio of the computational domain and the tangent of the inclination. In our 
models, this geometrical effect leads to another maximum in extracted jet widths
whose position moves to smaller distances with decreasing inclination (Fig. 
\ref{Fig_inclination_SC1c}) and has to be corrected when comparing our models 
to observations. We used this test case with constant emission for quantifying 
the area in which our extracted jet widths are not affected by this effect.

\section{Best-fit models} \label{sec_bestfit}

In paper I, we found best-fit models for the jets in the observed sample. Note 
that we always ignored there the first maximum in the synthetic jet widths and 
focussed on larger distances from the source. Now we can include the effects of 
inclination, use the position of the maxima in our synthetical jet width 
variations for estimating the inclination and compare the derived inclination 
with values from the literature.

\subsection{DG Tau}

\begin{figure}[!htb]
  \centering
  \includegraphics[width=\columnwidth]{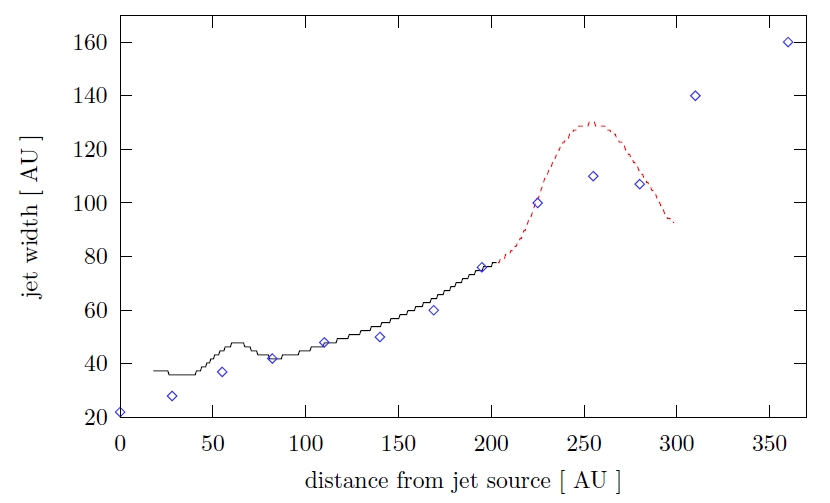}
  \caption{Best-fit model for DG Tau: model SC1a, run (500,1000,0.8) and 
    inclination of 40$^\circ$. The red part is not reliable due to the 
    geometric effect discussed in Sec. \ref{sec_variations}.}
  \label{Fig_DG_Tau}
\end{figure}

The observed mass of DG Tau (diamonds in Fig. \ref{Fig_observations}) is 
0.67 $M_\odot$ \citep{HEG95}, therefore we have to focus on the runs 
(500,1000,0.8), and perhaps also runs (500,600,0.5) and (500,1000,0.5). The 
best-fit model is between ADO and SC1a, thus the truncation radius is larger 
than 0.22 AU. DG Tau can be quite well reproduced with the model SC1a, the run 
(500,1000,0.8) and an inclination of 40$^\circ$ (Fig. \ref{Fig_DG_Tau}). This 
inclination is in excellent agreement with the literature values of 
32--52$^\circ$ \citep{EiM98,BRM02,PKH03}.

\subsection{CW Tau}

\begin{figure}[!htb]
  \centering
  \includegraphics[width=\columnwidth]{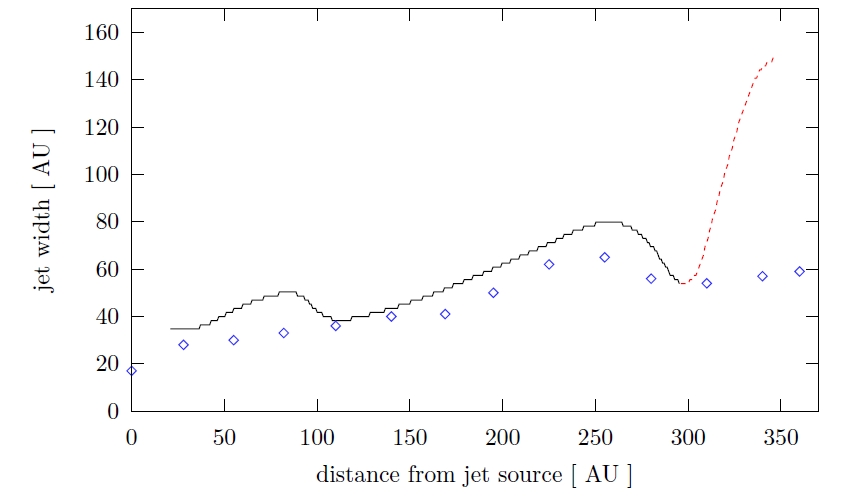}
  \caption{Best-fit model for CW Tau: model SC1c, run (500,1000,0.8) and 
    inclination of 60$^\circ$. The red part is not reliable due to the 
    geometric effect discussed in Sec. \ref{sec_variations}.}
  \label{Fig_CW_Tau}
\end{figure}

The mass of CW Tau (squares) is the highest in our sample, 1.03 $M_\odot$ 
\citep{HEG95}. Using the runs (500,1000,0.8), the best-fit model was SC1b or 
SC1c. The truncation radius is thus between 0.25 -- 0.3 AU. CW Tau may be 
best-modeled with simulation SC1c, run (500,1000,0.8) and an inclination of 
60$^\circ$ (Fig. \ref{Fig_CW_Tau}). The literature value, however, is about 
41$^\circ$ \citep{CBR07}.

\subsection{RW Aur}

The measured mass of RW Aur (circles) is 0.85 $M_\odot$ \citep{HEG95}, 
thus we had to focus on runs (500,1000,0.8). We needed a very high degree of 
truncation as in models SC1e-g, thus a truncation radius of the order of 0.04 
AU, which seems to be unphysical. For these models, however, the influence of 
inclination is only marginal; we find literature values of 42$^\circ$ 
\citep{LCH97}.

\subsection{HN Tau and UZ Tau E}

HN Tau (plus signs) has a mass of 0.72 $M_\odot$ \citep{HEG95}, thus 
again the runs (500,1000,0.8) were favored. When we ignore the first maximum, 
we can interpolate the jet shape at larger distances and find a best-fit model 
between ADO and SC1a, thus the truncation radius is again larger than 0.34 AU. 
However, this result is highly uncertain and this procedure contradicts our 
present approach of taking the maxima into account. 

UZ Tau E (crosses) has the lowest mass in our sample, only 0.18 $M_\odot$ 
\citep{HEG95}, thus we used the runs (500,600,0.2). Again we have to 
interpolate the jet width from larger distances and could in principle choose 
model SC1a as best-fit model. The truncation radius would be about 0.26 AU.

Over all, we cannot find best-fit models, since our numerical resolution very 
close to the jet source is not sufficient enough; the inclination of HN Tau is 
not known \citep{HEP04}, for that of UZ Tau E we find literature values of 
60--70$^\circ$ \citep{JKM96,PSM02}.

\section{Summary and conclusions}

In paper I, we showed as a proof of concept that jet widths derived from 
numerical simulations extending analytical MHD jet formation models can be very 
helpful for understanding recently observed jet widths from observations with 
adaptive optics and space telescopes. Here we investigated qualitatively the 
influence of different inclinations. 

We examined the origin of variations in the extracted jet widths and identified
physical and geometrical effects. We showed in which range our results are 
reliable and not affected by numerical artefacts.

We found that the distance of the first maximum from the jet source increases 
for increasing mass of the central object and decreasing velocity of the jet. It
also increases for increasing degree of truncation. In several models, a second 
maximum appears whose distance decreases for increasing degree of truncation. 
The most noticeable effect of a decreasing inclination is a change in the 
position of the maxima in the synthetic jet widths. In all cases, the distance 
of the first maximum from the jet source decreases for decreasing inclination. 
The lower the inclination, the higher is also the number of maxima in the jet 
width.

We compared our synthetic jet widths with observations of our sample including
the position of maxima in the jet width variations as diagnostics for the
inclination. Only for DG Tau and CW Tau, we could unambiguously find 
models which fitted the observed jet widths. For the former, the derived 
inclination is perfectly consistent with values found in the literature; for the
latter our derived inclination is higher than commonly measured (60$^\circ$ 
compared to 41$^\circ$).

In a future work, we have to refine our grid of models for reproducing the 
observed jet widths and their variations even more accurate. These models have 
to be done with higher numerical resolution in order to give significant results
in terms of synthetic emission maps even within a few AU to the jet source. 
Furthermore we have to extend the dimensions of our computational domain in 
order to avoid the described geometrical artefacts.

\begin{acknowledgements}
The authors would like to thank the referee for suggestions and comments which
improved this paper.
\end{acknowledgements}

\clearpage

\appendix

\section{Extracted line widths for all models and runs} \label{sec_all_lw}

Here we present the extracted line widths derived from synthetic [OI] 
$\lambda$6300 images for all models, runs and inclinations for the sake of 
completeness. 

\begin{figure}[!b]
  \centering
  \includegraphics[width=0.45\textwidth]{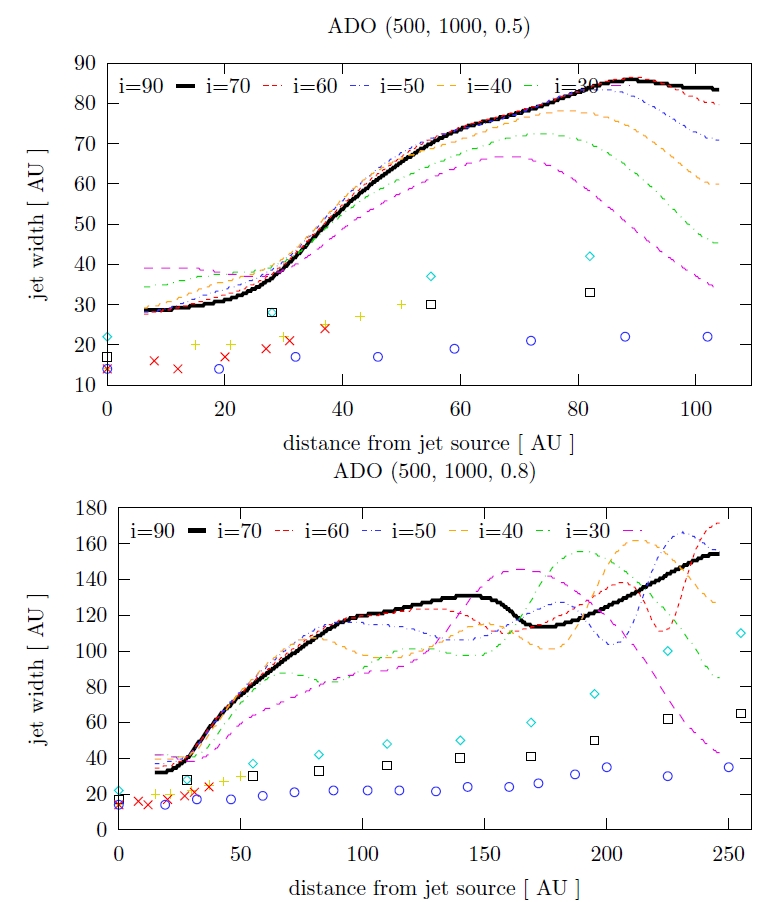}
  \includegraphics[width=0.45\textwidth]{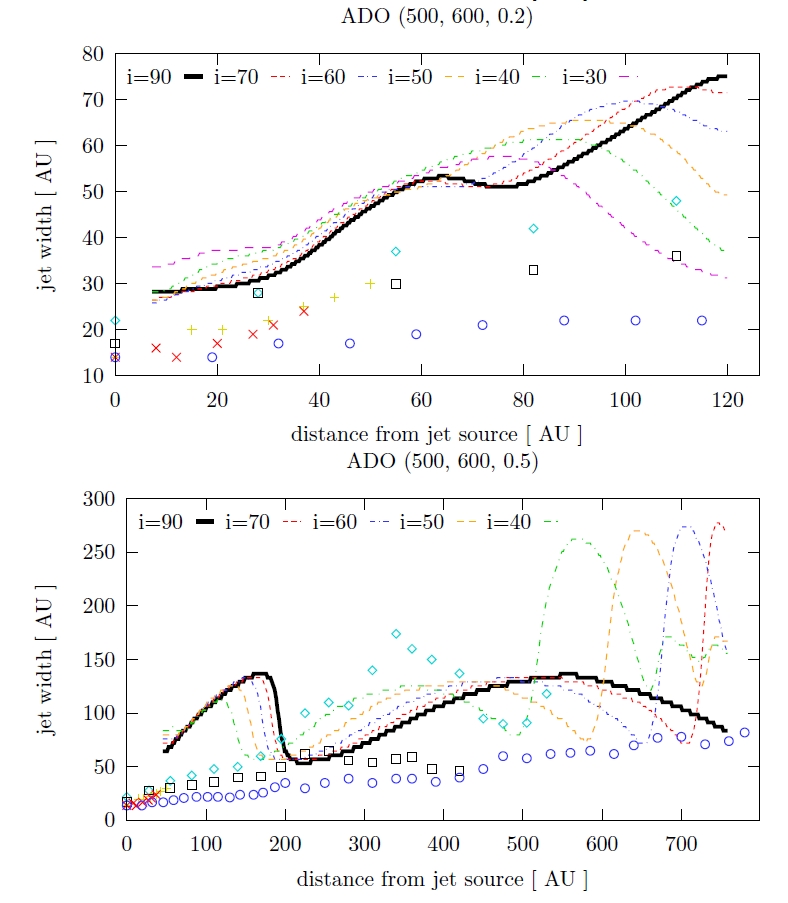}
  \caption{Jet widths in AU derived from synthetic [OI] images as a 
    function of distance from the source in model ADO and for runs 
    (500,600,0.2), (500,600,0.5), (500,1000,0.5), (500,1000,0.8); overlaid are 
    the data points of Fig. \ref{Fig_observations}.}
  \label{Fig_all_inclination_ADO}
\end{figure}

\begin{figure}[!b]
  \centering
  \includegraphics[width=0.45\textwidth]{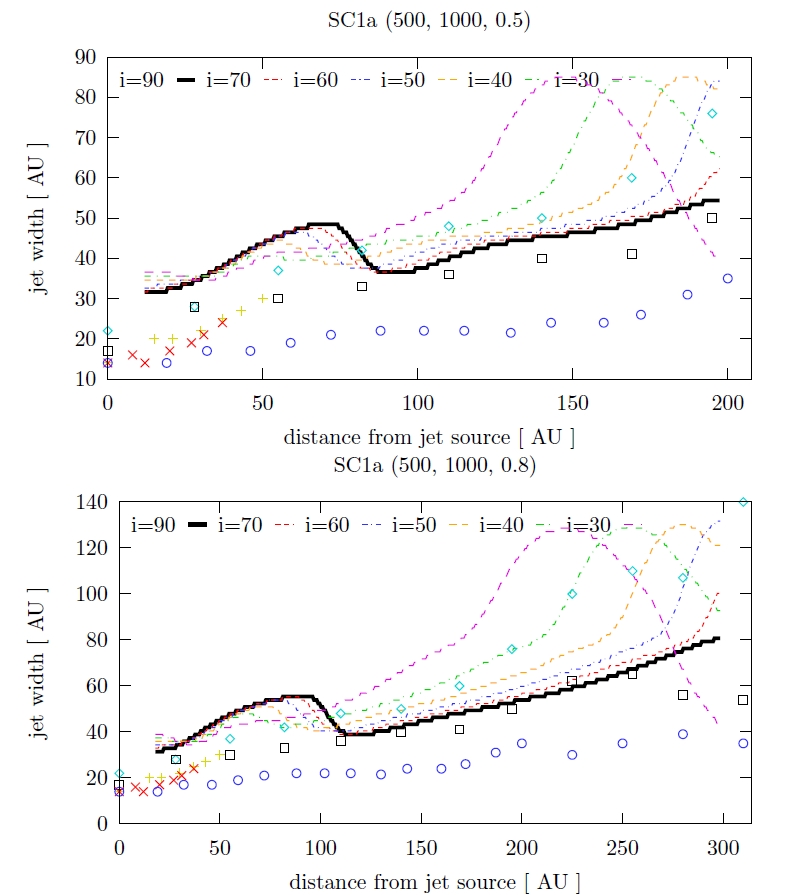}
  \includegraphics[width=0.45\textwidth]{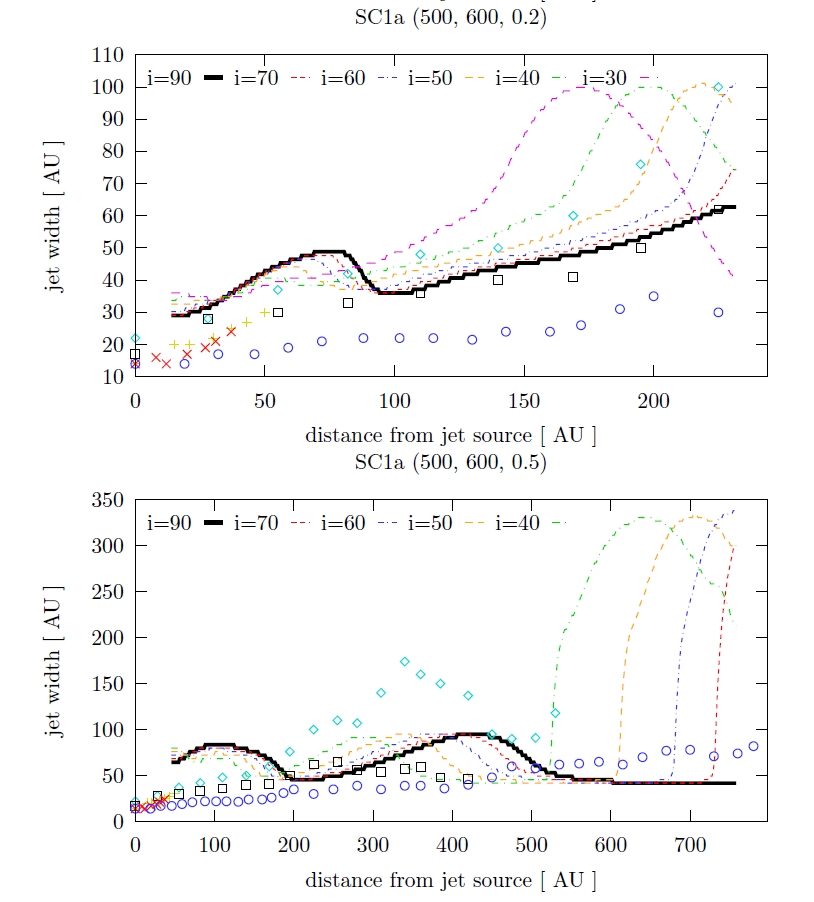}
  \caption{same as Fig. \ref{Fig_all_inclination_ADO}, but in model SC1a.}
  \label{Fig_all_inclination_SC1}
\end{figure}

\begin{figure}[!htb]
  \centering
  \includegraphics[width=0.45\textwidth]{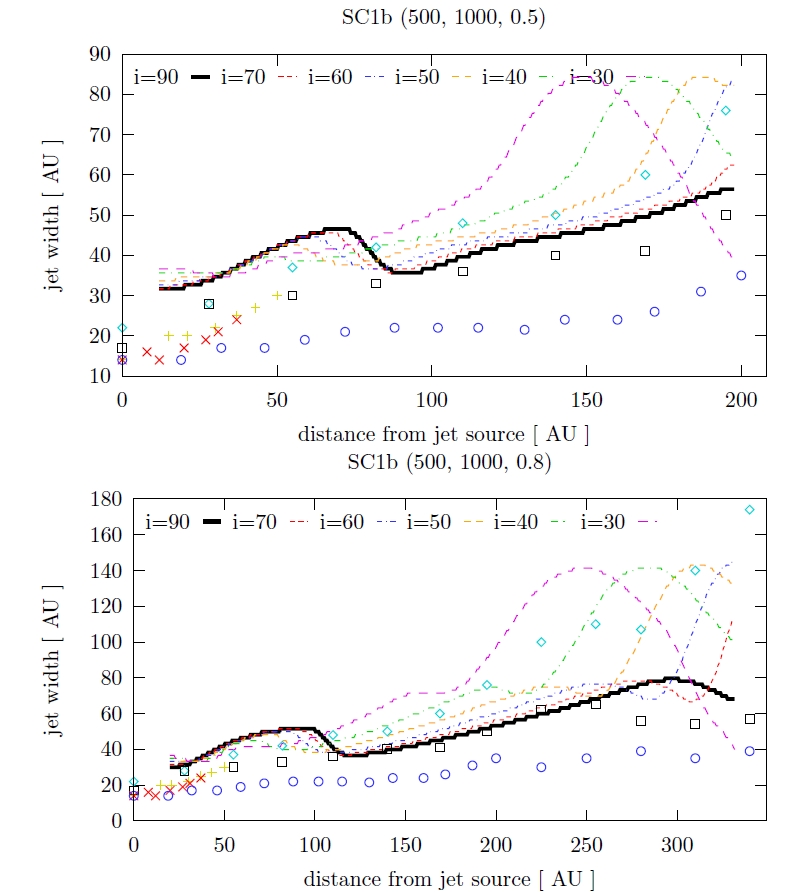}
  \includegraphics[width=0.45\textwidth]{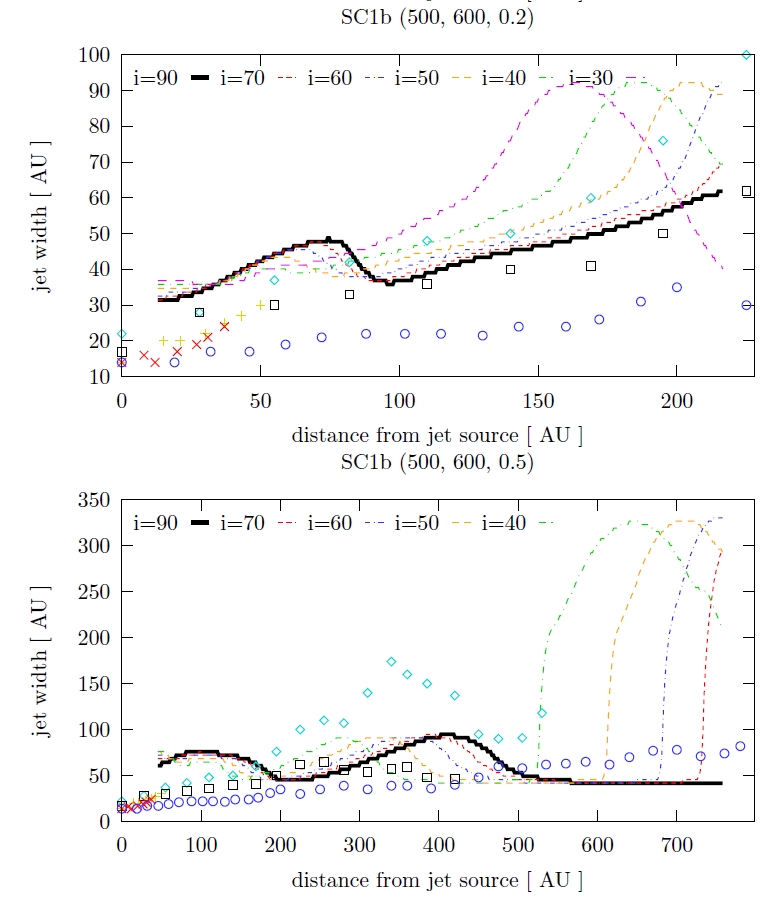}
  \caption{same as Fig. \ref{Fig_all_inclination_ADO}, but in model SC1b.}
  \label{Fig_all_inclination_SC1b}
\end{figure}

\begin{figure}[!htb]
  \centering
  \includegraphics[width=0.45\textwidth]{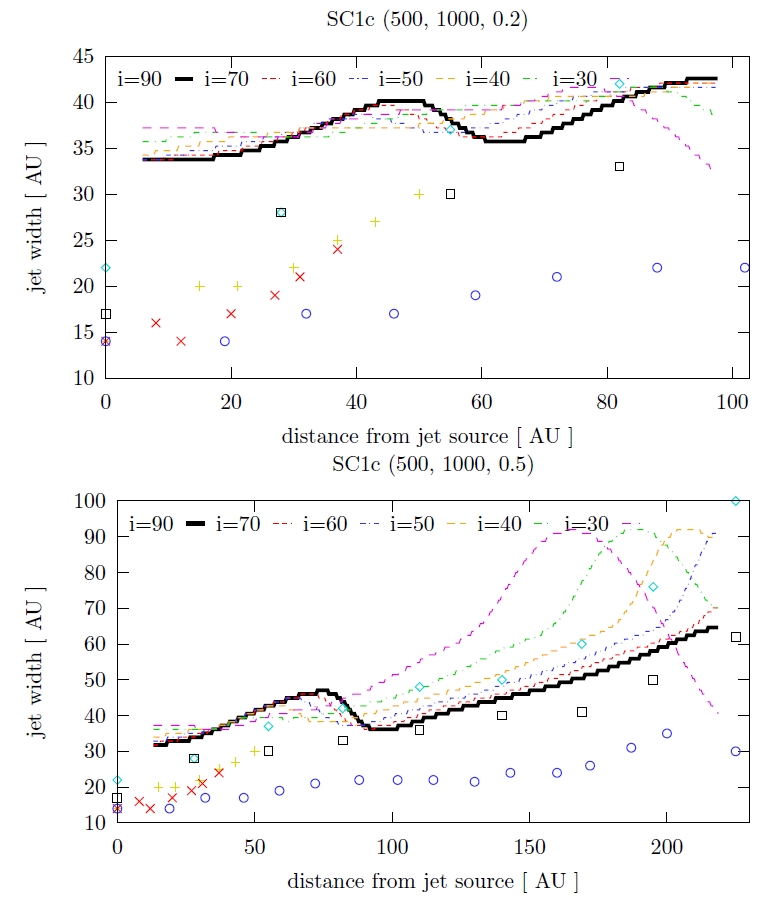}
  \includegraphics[width=0.45\textwidth]{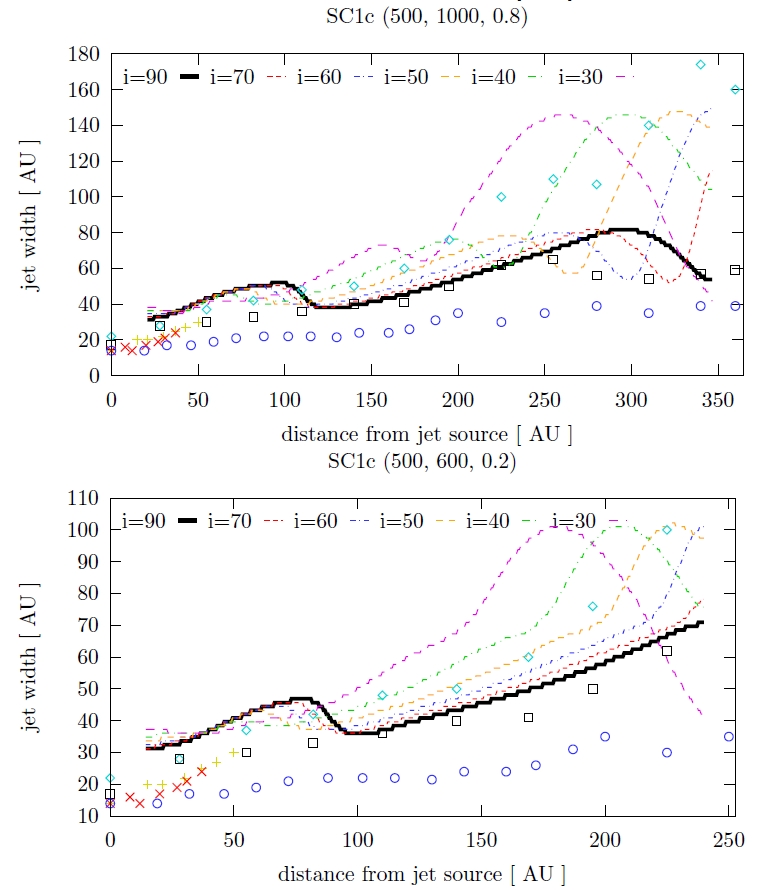}
  \includegraphics[width=0.45\textwidth]{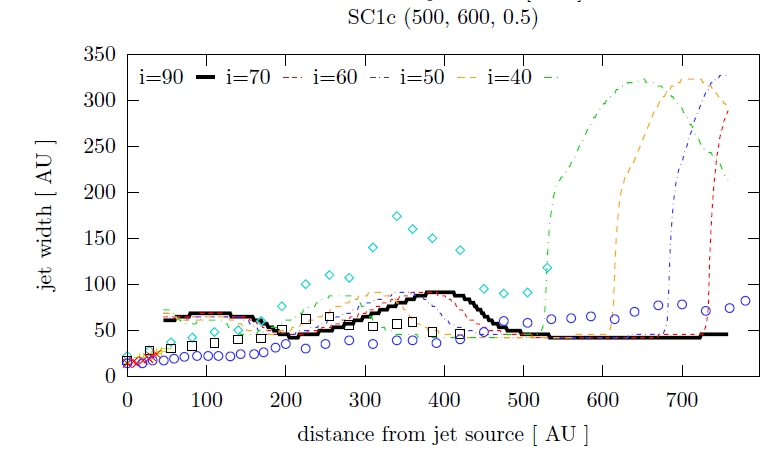}
  \caption{same as Fig. \ref{Fig_all_inclination_ADO}, but in model SC1c.}
  \label{Fig_all_inclination_SC1c}
\end{figure}

\begin{figure}[!htb]
  \centering
  \includegraphics[width=0.45\textwidth]{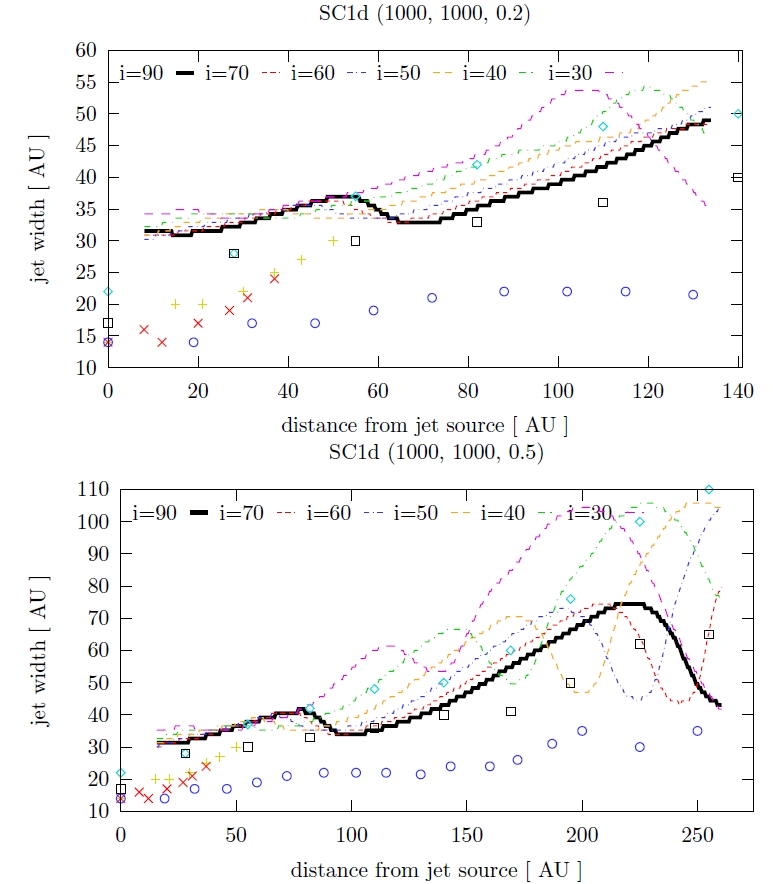}
  \includegraphics[width=0.45\textwidth]{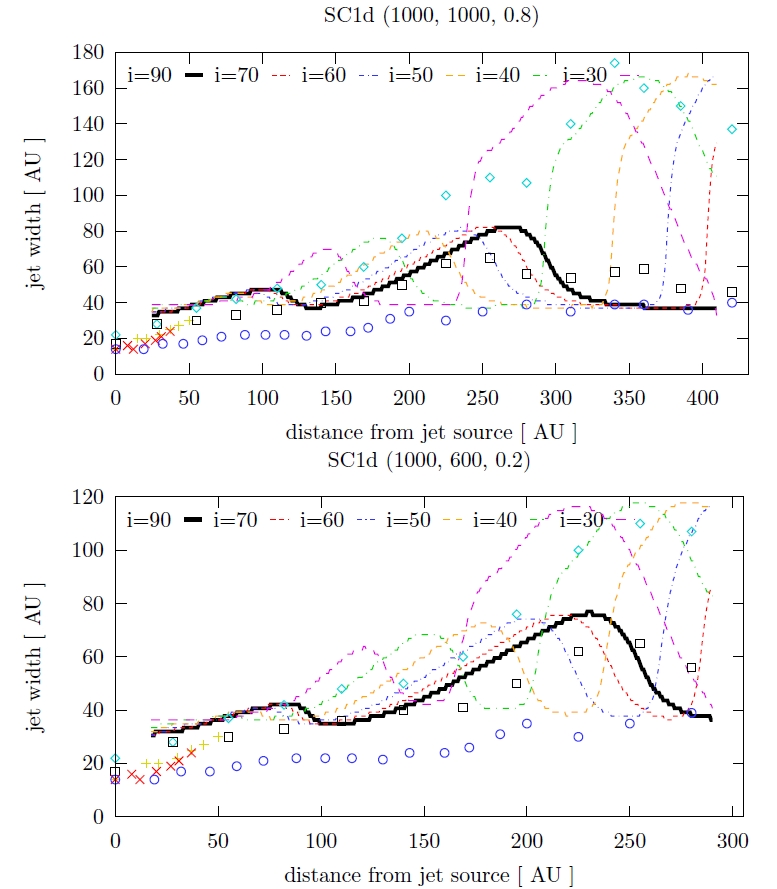}
  \includegraphics[width=0.45\textwidth]{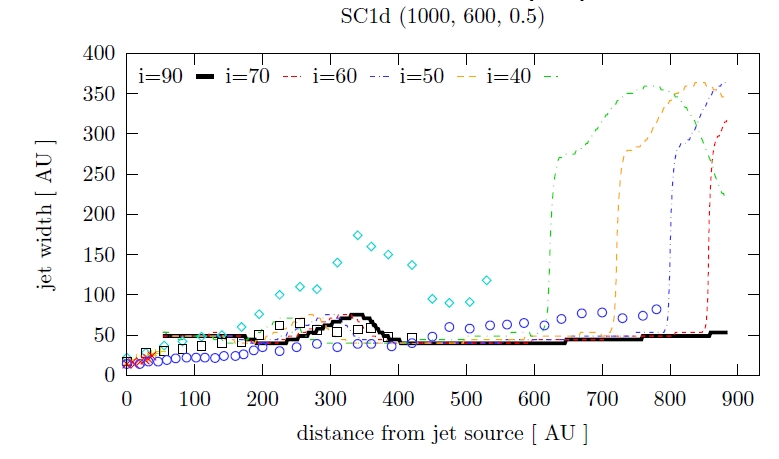}
  \caption{same as Fig. \ref{Fig_all_inclination_ADO}, but in model SC1d.}
  \label{Fig_all_inclination_SC1d}
\end{figure}

\begin{figure}[!htb]
  \centering
  \includegraphics[width=0.45\textwidth]{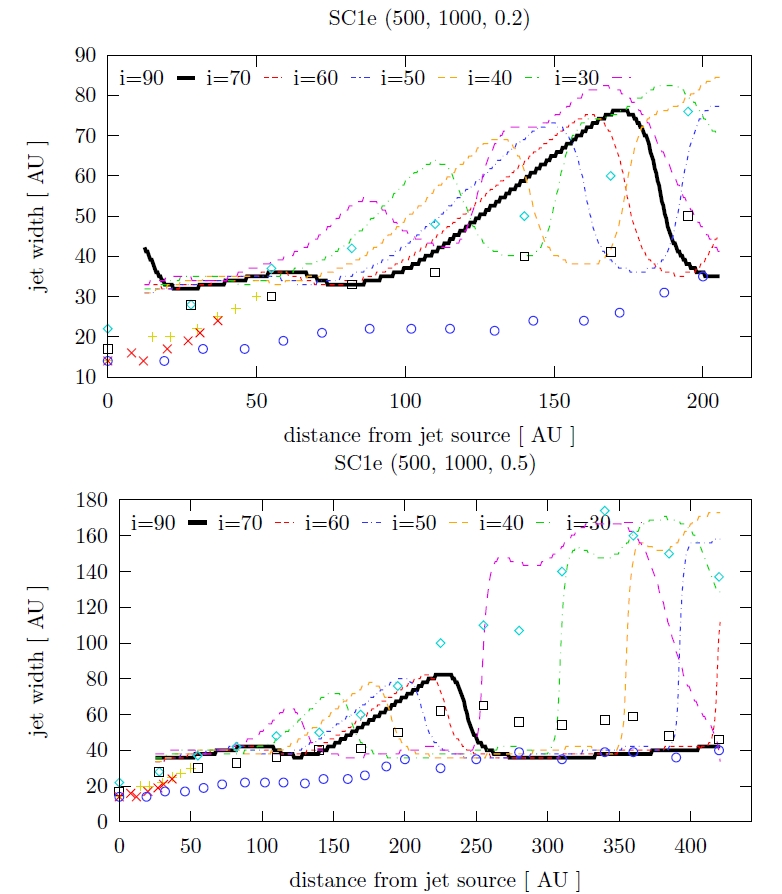}
  \includegraphics[width=0.45\textwidth]{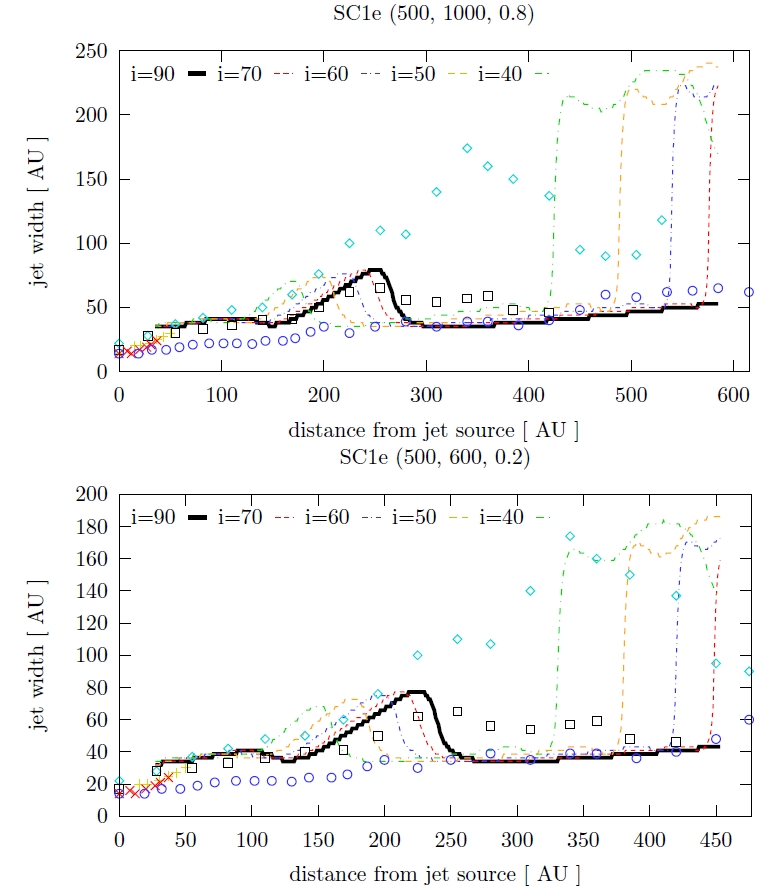}
  \includegraphics[width=0.45\textwidth]{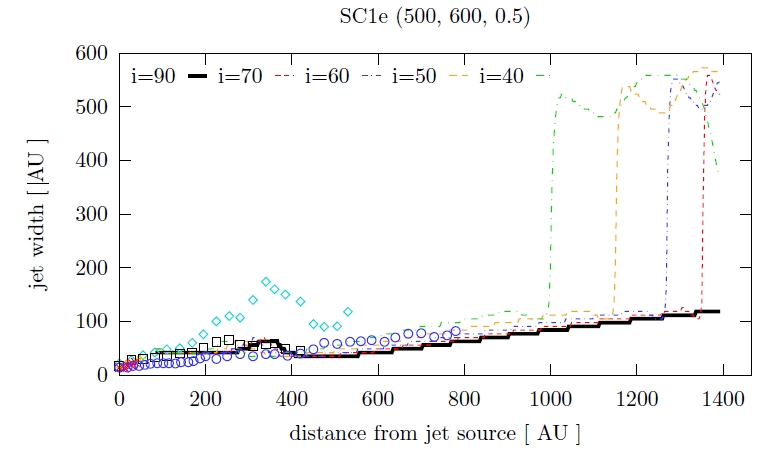}
  \caption{same as Fig. \ref{Fig_all_inclination_ADO}, but in model SC1e.}
  \label{Fig_all_inclination_SC1e}
\end{figure}

\begin{figure}[!htb]
  \centering
  \includegraphics[width=0.45\textwidth]{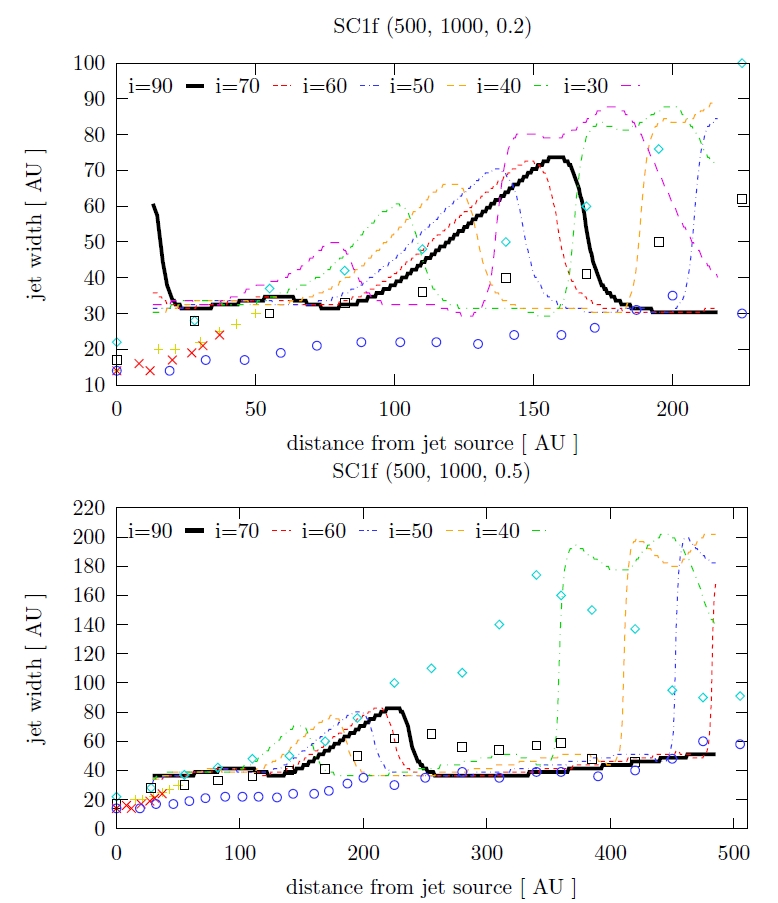}
  \includegraphics[width=0.45\textwidth]{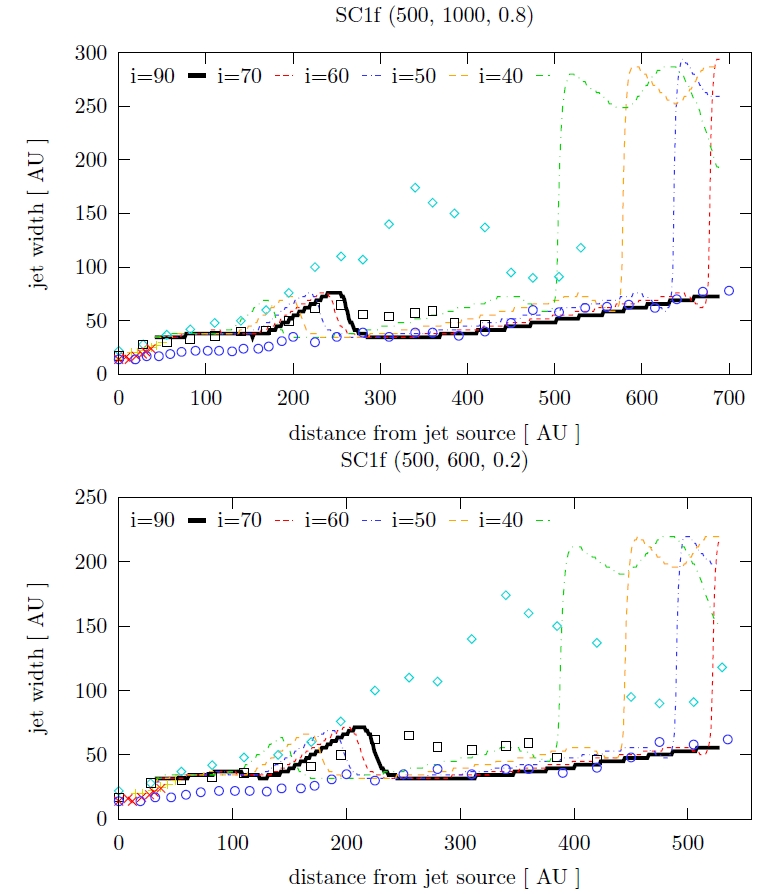}
  \includegraphics[width=0.45\textwidth]{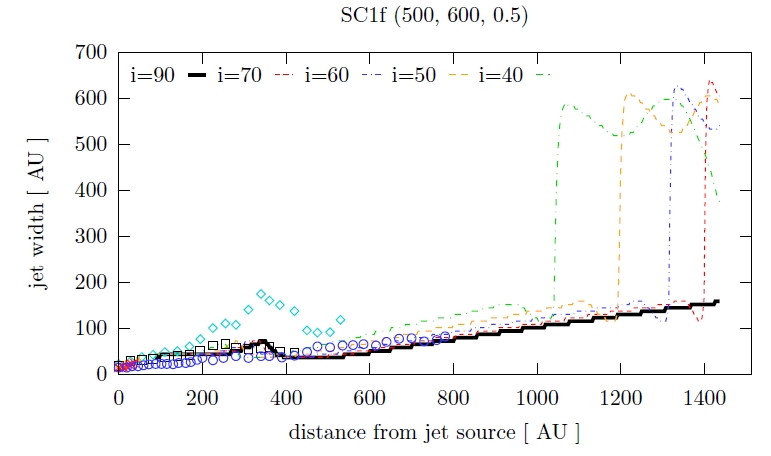}
  \caption{same as Fig. \ref{Fig_all_inclination_ADO}, but in model SC1f.}
  \label{Fig_all_inclination_SC1f}
\end{figure}

\begin{figure}[!htb]
  \centering
  \includegraphics[width=0.45\textwidth]{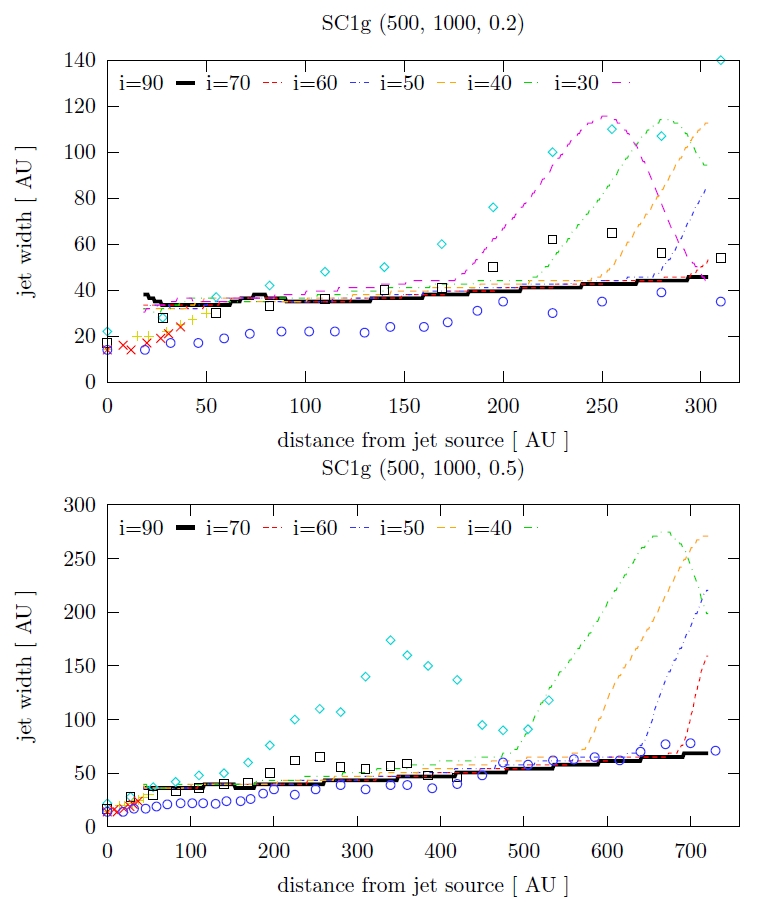}
  \includegraphics[width=0.45\textwidth]{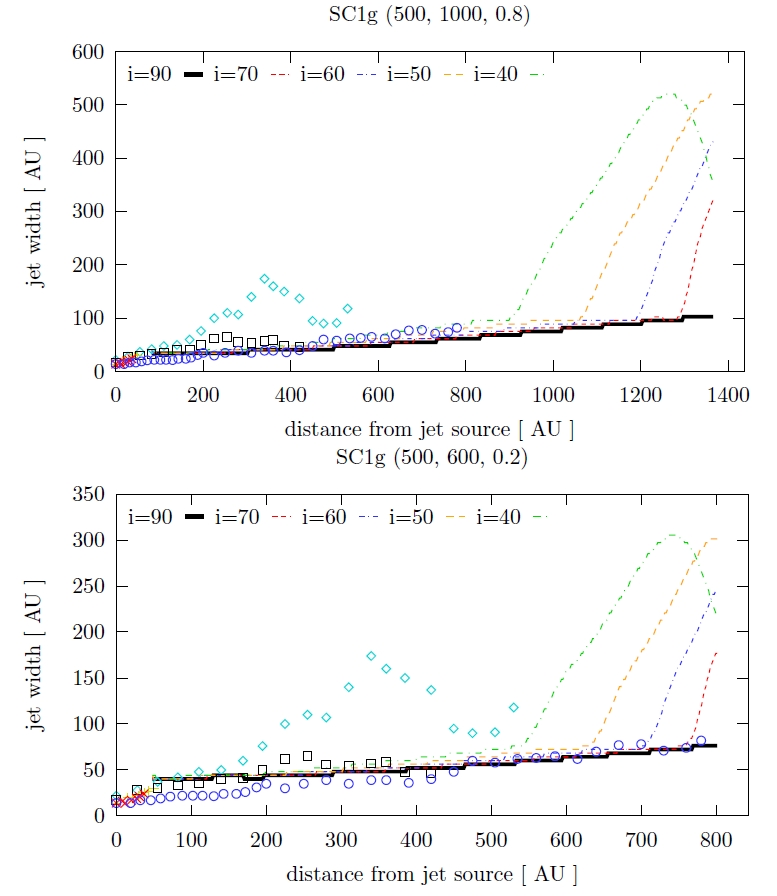}
  \caption{same as Fig. \ref{Fig_all_inclination_ADO}, but in model SC1g.}
  \label{Fig_all_inclination_SC1g}
\end{figure}

\clearpage

\section{Velocities of the jet derived from the synthetic 
position-velocity diagrams} \label{sec_pv_vel}

\begin{table}[!htb]
\centering
\begin{tabular}{c c c c | c}
\hline\hline
model & $v_{\rm jet}$ & $M$ & inclination & $v_{\rm jet, PV}$ \\
\hline
ADO & 600 & 0.2 & 10 & 140 \\
ADO & 600 & 0.2 & 20 & 130 \\
ADO & 600 & 0.2 & 30 & 120 \\
ADO & 600 & 0.2 & 40 &  73 \\
ADO & 600 & 0.2 & 50 &  68 \\
ADO & 600 & 0.2 & 60 &  58 \\
ADO & 600 & 0.2 & 70 &  47 \\
ADO & 600 & 0.2 & 80 &  32 \\
ADO & 600 & 0.2 & 90 &   0 \\
\hline
ADO & 600 & 0.5 & 10 & 527 \\
ADO & 600 & 0.5 & 20 & 396 \\
ADO & 600 & 0.5 & 30 & 375 \\
ADO & 600 & 0.5 & 40 & 302 \\
ADO & 600 & 0.5 & 50 & 172 \\
ADO & 600 & 0.5 & 60 &  99 \\
ADO & 600 & 0.5 & 70 &  52 \\
ADO & 600 & 0.5 & 80 &  16 \\
ADO & 600 & 0.5 & 90 &   0 \\
\hline
ADO & 1000 & 0.5 & 10 & 120 \\
ADO & 1000 & 0.5 & 20 &  99 \\
ADO & 1000 & 0.5 & 30 &  99 \\
ADO & 1000 & 0.5 & 40 &  89 \\
ADO & 1000 & 0.5 & 50 &  99 \\
ADO & 1000 & 0.5 & 60 &  94 \\
ADO & 1000 & 0.5 & 70 &  83 \\
ADO & 1000 & 0.5 & 80 &  37 \\
ADO & 1000 & 0.5 & 90 &   0 \\
\hline
ADO & 1000 & 0.8 & 10 & 141 \\
ADO & 1000 & 0.8 & 20 & 141 \\
ADO & 1000 & 0.8 & 30 & 130 \\
ADO & 1000 & 0.8 & 40 & 130 \\
ADO & 1000 & 0.8 & 50 & 130 \\
ADO & 1000 & 0.8 & 60 & 109 \\
ADO & 1000 & 0.8 & 70 &  78 \\
ADO & 1000 & 0.8 & 80 &  37 \\
ADO & 1000 & 0.8 & 90 &   0 \\
\hline
\hline
\end{tabular}
\end{table}
\begin{table}[!htb]
\centering
\begin{tabular}{c c c c | c}
\hline\hline
model & $v_{\rm jet}$ & $M$ & inclination & $v_{\rm jet, PV}$ \\
\hline
SC1a & 600 & 0.2 & 10 & 193 \\
SC1a & 600 & 0.2 & 20 & 198 \\
SC1a & 600 & 0.2 & 30 & 167 \\
SC1a & 600 & 0.2 & 40 & 115 \\
SC1a & 600 & 0.2 & 50 &  58 \\
SC1a & 600 & 0.2 & 60 &  42 \\
SC1a & 600 & 0.2 & 70 &  63 \\
SC1a & 600 & 0.2 & 80 &  42 \\
SC1a & 600 & 0.2 & 90 &   0 \\
\hline
SC1a & 600 & 0.5 & 10 & 760 \\
SC1a & 600 & 0.5 & 20 & 500 \\
SC1a & 600 & 0.5 & 30 & 271 \\
SC1a & 600 & 0.5 & 40 &  89 \\
SC1a & 600 & 0.5 & 50 &  37 \\
SC1a & 600 & 0.5 & 60 &  99 \\
SC1a & 600 & 0.5 & 70 & 104 \\
SC1a & 600 & 0.5 & 80 &  73 \\
SC1a & 600 & 0.5 & 90 &   0 \\
\hline
SC1a & 1000 & 0.5 & 10 & 230 \\
SC1a & 1000 & 0.5 & 20 & 276 \\
SC1a & 1000 & 0.5 & 30 & 260 \\
SC1a & 1000 & 0.5 & 40 & 183 \\
SC1a & 1000 & 0.5 & 50 & 109 \\
SC1a & 1000 & 0.5 & 60 &  68 \\
SC1a & 1000 & 0.5 & 70 &  94 \\
SC1a & 1000 & 0.5 & 80 &  32 \\
SC1a & 1000 & 0.5 & 90 &   0 \\
\hline
SC1a & 1000 & 0.8 & 10 & 433 \\
SC1a & 1000 & 0.8 & 20 & 520 \\
SC1a & 1000 & 0.8 & 30 & 401 \\
SC1a & 1000 & 0.8 & 40 & 220 \\
SC1a & 1000 & 0.8 & 50 &  63 \\
SC1a & 1000 & 0.8 & 60 &  94 \\
SC1a & 1000 & 0.8 & 70 &  58 \\
SC1a & 1000 & 0.8 & 80 &  47 \\
SC1a & 1000 & 0.8 & 90 &   0 \\
\hline
\hline
\end{tabular}
\end{table}
\begin{table}[!htb]
\centering
\begin{tabular}{c c c c | c}
\hline\hline
model & $v_{\rm jet}$ & $M$ & inclination & $v_{\rm jet, PV}$ \\
\hline
SC1b & 600 & 0.2 & 10 & 198 \\
SC1b & 600 & 0.2 & 20 & 198 \\
SC1b & 600 & 0.2 & 30 & 183 \\
SC1b & 600 & 0.2 & 40 & 115 \\
SC1b & 600 & 0.2 & 50 &  68 \\
SC1b & 600 & 0.2 & 60 &  32 \\
SC1b & 600 & 0.2 & 70 &  78 \\
SC1b & 600 & 0.2 & 80 &  52 \\
SC1b & 600 & 0.2 & 90 &   0 \\
\hline
SC1b & 600 & 0.5 & 10 & 780 \\
SC1b & 600 & 0.5 & 20 & 505 \\
SC1b & 600 & 0.5 & 30 & 276 \\
SC1b & 600 & 0.5 & 40 &  89 \\
SC1b & 600 & 0.5 & 50 &  42 \\
SC1b & 600 & 0.5 & 60 &  94 \\
SC1b & 600 & 0.5 & 70 & 109 \\
SC1b & 600 & 0.5 & 80 &  68 \\
SC1b & 600 & 0.5 & 90 &   0 \\
\hline
SC1b & 1000 & 0.5 & 10 & 328 \\
SC1b & 1000 & 0.5 & 20 & 286 \\
SC1b & 1000 & 0.5 & 30 & 244 \\
SC1b & 1000 & 0.5 & 40 & 177 \\
SC1b & 1000 & 0.5 & 50 & 120 \\
SC1b & 1000 & 0.5 & 60 &  73 \\
SC1b & 1000 & 0.5 & 70 &  68 \\
SC1b & 1000 & 0.5 & 80 &  63 \\
SC1b & 1000 & 0.5 & 90 &   0 \\
\hline
SC1b & 1000 & 0.8 & 10 & 469 \\
SC1b & 1000 & 0.8 & 20 & 497 \\
SC1b & 1000 & 0.8 & 30 & 406 \\
SC1b & 1000 & 0.8 & 40 & 198 \\
SC1b & 1000 & 0.8 & 50 &  78 \\
SC1b & 1000 & 0.8 & 60 &  68 \\
SC1b & 1000 & 0.8 & 70 &  63 \\
SC1b & 1000 & 0.8 & 80 &  36 \\
SC1b & 1000 & 0.8 & 90 &   0 \\
\hline
\hline
\end{tabular}
\end{table}
\begin{table}[!htb]
\centering
\begin{tabular}{c c c c | c}
\hline\hline
model & $v_{\rm jet}$ & $M$ & inclination & $v_{\rm jet, PV}$ \\
\hline
SC1c & 600 & 0.2 & 10 & 236 \\
SC1c & 600 & 0.2 & 20 & 224 \\
SC1c & 600 & 0.2 & 30 & 177 \\
SC1c & 600 & 0.2 & 40 & 121 \\
SC1c & 600 & 0.2 & 50 &  68 \\
SC1c & 600 & 0.2 & 60 &  32 \\
SC1c & 600 & 0.2 & 70 &  73 \\
SC1c & 600 & 0.2 & 80 &  52 \\
SC1c & 600 & 0.2 & 90 &   0 \\
\hline
SC1c & 600 & 0.5 & 10 & 770 \\
SC1c & 600 & 0.5 & 20 & 521 \\
SC1c & 600 & 0.5 & 30 & 286 \\
SC1c & 600 & 0.5 & 40 & 110 \\
SC1c & 600 & 0.5 & 50 &  33 \\
SC1c & 600 & 0.5 & 60 &  94 \\
SC1c & 600 & 0.5 & 70 & 104 \\
SC1c & 600 & 0.5 & 80 &  68 \\
SC1c & 600 & 0.5 & 90 &   0 \\
\hline
SC1c & 1000 & 0.5 & 10 & 375 \\
SC1c & 1000 & 0.5 & 20 & 307 \\
SC1c & 1000 & 0.5 & 30 & 259 \\
SC1c & 1000 & 0.5 & 40 & 193 \\
SC1c & 1000 & 0.5 & 50 & 115 \\
SC1c & 1000 & 0.5 & 60 &  68 \\
SC1c & 1000 & 0.5 & 70 &  94 \\
SC1c & 1000 & 0.5 & 80 &  58 \\
SC1c & 1000 & 0.5 & 90 &   0 \\
\hline
SC1c & 1000 & 0.8 & 10 & 479 \\
SC1c & 1000 & 0.8 & 20 & 479 \\
SC1c & 1000 & 0.8 & 30 & 443 \\
SC1c & 1000 & 0.8 & 40 & 208 \\
SC1c & 1000 & 0.8 & 50 &  74 \\
SC1c & 1000 & 0.8 & 60 &  68 \\
SC1c & 1000 & 0.8 & 70 &  84 \\
SC1c & 1000 & 0.8 & 80 &  73 \\
SC1c & 1000 & 0.8 & 90 &   0 \\
\hline
\hline
\end{tabular}
\end{table}
\begin{table}[!htb]
\centering
\begin{tabular}{c c c c | c}
\hline\hline
model & $v_{\rm jet}$ & $M$ & inclination & $v_{\rm jet, PV}$ \\
\hline
SC1d & 600 & 0.2 & 10 & 339 \\
SC1d & 600 & 0.2 & 20 & 261 \\
SC1d & 600 & 0.2 & 30 & 183 \\
SC1d & 600 & 0.2 & 40 & 183 \\
SC1d & 600 & 0.2 & 50 &  99 \\
SC1d & 600 & 0.2 & 60 &  43 \\
SC1d & 600 & 0.2 & 70 &  67 \\
SC1d & 600 & 0.2 & 80 &  52 \\
SC1d & 600 & 0.2 & 90 &   0 \\
\hline
SC1d & 600 & 0.5 & 10 & 744 \\
SC1d & 600 & 0.5 & 20 & 516 \\
SC1d & 600 & 0.5 & 30 & 307 \\
SC1d & 600 & 0.5 & 40 & 135 \\
SC1d & 600 & 0.5 & 50 &  26 \\
SC1d & 600 & 0.5 & 60 &  84 \\
SC1d & 600 & 0.5 & 70 &  89 \\
SC1d & 600 & 0.5 & 80 &  58 \\
SC1d & 600 & 0.5 & 90 &   0 \\
\hline
SC1d & 1000 & 0.5 & 10 & 485 \\
SC1d & 1000 & 0.5 & 20 & 401 \\
SC1d & 1000 & 0.5 & 30 & 260 \\
SC1d & 1000 & 0.5 & 40 & 188 \\
SC1d & 1000 & 0.5 & 50 & 125 \\
SC1d & 1000 & 0.5 & 60 &  78 \\
SC1d & 1000 & 0.5 & 70 &  83 \\
SC1d & 1000 & 0.5 & 80 &  47 \\
SC1d & 1000 & 0.5 & 90 &   0 \\
\hline
SC1d & 1000 & 0.8 & 10 & 594 \\
SC1d & 1000 & 0.8 & 20 & 453 \\
SC1d & 1000 & 0.8 & 30 & 432 \\
SC1d & 1000 & 0.8 & 40 & 250 \\
SC1d & 1000 & 0.8 & 50 & 104 \\
SC1d & 1000 & 0.8 & 60 & 115 \\
SC1d & 1000 & 0.8 & 70 & 130 \\
SC1d & 1000 & 0.8 & 80 &  94 \\
SC1d & 1000 & 0.8 & 90 &   0 \\
\hline
\hline
\end{tabular}
\end{table}


\begin{thebibliography}{}
\bibitem[Anderson et al.(2003)]{ALK03}
Anderson, J. M., Li, Z.-Y., Krasnopolsky, R., Blandford, R. D. 2003, ApJ, 590, 
L107
\bibitem[Bacciotti \& Eisl\"offel(1999)]{BE99}
Bacciotti, F., Eisl\"offel, J. 1999, A \& A, 342, 717
\bibitem[Bacciotti et al.(2002)]{BRM02}
Bacciotti, F., Ray, T. P., Mundt, R., et al. 2002, ApJ, 576, 222
\bibitem[Blandford \& Payne(1982)]{BlP82}
Blandford, R. D., Payne, D. G. 1982, MNRAS, 199, 883
\bibitem[Cabrit et al.(1990)]{CES90} 
Cabrit, S., Edwards, S., Strom, S. E., Strom, K. M. 1990, ApJ, 354, 687
\bibitem[Casse \& Ferreira(2000)]{CaF00}
Casse, F., Ferreira, J. 2000, A \& A, 353, 1115
\bibitem[Coffey et al.(2007)]{CBR07}
Coffey, D., Bacciotti, F., Ray, T. P., et al. 2007, ApJ, 663, 350
\bibitem[Combet \& Ferreira(2008)]{CoF08}
Combet, C., Ferreira, J. 2008, A \& A, 479, 481
\bibitem[Dougados et al. (2000)]{DCL00}
Dougados, C., Cabrit, S., Lavalley, C., Menard, F. 2000, A\&A, 357, L61
\bibitem[Dougados et al. (2004)]{DCF04}
Dougados, C., Cabrit, S., Ferreira, J., Pesenti, N., Garcia, P., O'Brien, D. 
2004, Ap \& SS, 293, 45
\bibitem[Dougados(2008)]{Dou08}
Dougados, C. 2008, in: ``Jets from Young Stars II: Clues from High Angular 
Resolution Observations'', Lecture Notes in Physics, Vol. 742, F. Bacciotti, E. 
Whelan, L. Testi (Eds.), Springer-Verlag Berlin Heidelberg
\bibitem[Eisl\"offel \& Mundt(1998)]{EiM98}
Eisl\"offel, J., Mundt, R. 1998, AJ, 115, 1554
\bibitem[Ferreira(1997)]{Fer97}
Ferreira, J. 1997, A \& A, 319, 340
\bibitem[Ferreira(2007)]{Fer07}
Ferreira, J. 2007, in: ``Jets from Young Stars: Models and Constraints'',
Lecture Notes in Physics, Vol. 723, J. Ferreira, C. Dougados, E. Whelan (Eds.), 
Springer-Verlag Berlin Heidelberg
\bibitem[Garcia et al.(2001)]{GCF01}
Garcia, P.J.V., Cabrit, S., Ferreira, J., Binette, L. 2001, A \& A, 377, 609
\bibitem[Gracia et al.(2006)]{GVT06}
Gracia, J., Vlahakis, N., Tsinganos, K. 2006, MNRAS, 367, 201, GVT06
\bibitem[Hartigan et al.(1995)]{HEG95} 
Hartigan, P., Edwards, S., Ghandour, L. 1995, ApJ, 452, 736
\bibitem[Hartigan et al.(2004)]{HEP04} 
Hartigan, P., Edwards, S., Pierson, R. 2004, ApJ, 609, 261
\bibitem[Hartmann(2009)]{Har09}
Hartmann, L. 2009, in ``Protostellar Jets in Context'', K. Tsinganos, T.P. Ray, 
M. Stute (Eds.), Springer-Verlag Berlin Heidelberg
\bibitem[Jensen et al.(1996)]{JKM96}
Jensen, E. L. N., Koerner, D. W., Mathieu, R. D. 1996, AJ, 111, 2431
\bibitem[Lavalley-Fouquet et al.(2000)]{LCD00}
Lavalley-Fouquet, C., Cabrit, S., Dougados, C. 2000, A \& A, 356, L41
\bibitem[Lay et al.(1997)]{LCH97}
Lay, O. P., Carlstrom, J. E., Hills, R. E. 1997, ApJ, 489, 917
\bibitem[Livio(2009)]{Liv09}
Livio, M. 2009, in ``Protostellar Jets in Context'', K. Tsinganos, T.P. Ray, 
M. Stute (Eds.), Springer-Verlag Berlin Heidelberg
\bibitem[Matsakos et al.(2008)]{MTV08}
Matsakos, T., Tsinganos, K., Vlahakis, N., Massaglia, S., Mignone, A., 
Trussoni, E. 2008, A \& A, 477, 521, M08
\bibitem[Matsakos et al.(2009)]{MMT09}
Matsakos, T., Massaglia, S., Trussoni, E., Tsinganos, K., Vlahakis, N., 
Sauty, C., Mignone, A. 2009, A \& A, 502, 217
\bibitem[Mignone et al.(2007)]{MBM07}
Mignone, A., Bodo, G., Massaglia, S., et al. 2007, ApJS, 170, 228
\bibitem[Najita et al. (2007)]{NCG07}
Najita, J. R., Carr, J. S., Glassgold, A. E., Valenti, J. A. 2007, in: 
``Protostars and Planets V'', Reipurth, B. Jewitt, D., Keil, K. (eds.), 
pp. 507--522. University of Arizona Press, Tucson (2007)
\bibitem[Prato et al.(2002)]{PSM02}
Prato, L., Simon, M., Mazeh, T., et al. 2002, ApJ, 579, L99
\bibitem[Pyo et al.(2003)]{PKH03}
Pyo, T.-S., Kobayashi, N., Hayashi, M., et al. 2003, ApJ, 590, 340
\bibitem[Ray et al.(1996)]{RMD96}
Ray, T. P., Mundt, R., Dyson, J. E., Falle, S. A. E. G., Raga, A. C. 1996, 
ApJ, 468, L103
\bibitem[Ray et al.(2007)]{RDB07}
Ray, T. P., Dougados, C., Bacciotti, F., et al. 2007, in: ``Protostars
and Planets V'', Reipurth, B. Jewitt, D., Keil, K. (eds.), pp. 231--244.
University of Arizona Press, Tucson (2007)
\bibitem[Stute et al.(2008)]{STV08}
Stute, M., Tsinganos, K., Vlahakis, N., Matsakos, T., Gracia, J. 2008, 
A \& A, 491, 339
\bibitem[Stute et al.(2010)]{SGT10}
Stute, M., Gracia, J., Tsinganos, K., Vlahakis, N. 2010, A \& A, 516, A6,
paper I
\bibitem[Vlahakis \& Tsinganos(1998)]{VlT98}
Vlahakis, N., Tsinganos, K. 1998, MNRAS, 298, 777
\bibitem[Vlahakis et al.(2000)]{VTS00}
Vlahakis, N., Tsinganos, K., Sauty, C., Trussoni, E. 2000, MNRAS, 318, 417, V00
\bibitem[Woitas et al.(2002)]{WRB02}
Woitas, J., Ray, T.P., Bacciotti, F., Davis, C.J., Eisl\"offel, J. 2002, 
ApJ, 580, 336
\bibitem[Zanni et al.(2007)]{ZFR07} 
Zanni, C., Ferrari, A., Rosner, R., Bodo, G., Massaglia, S. 2007, A \& A, 469, 
811
\end{thebibliography}
\end{document}